\documentclass[format=acmsmall, review=false, screen=true]{acmart}
\usepackage{booktabs} 

\setcopyright{acmcopyright}
\acmJournal{TKDD}
\acmYear{2020} \acmVolume{14} \acmNumber{5} \acmArticle{63} \acmMonth{8} \acmPrice{15.00}
\acmDOI{10.1145/3397191}

\usepackage{array}
\newcolumntype{P}[1]{>{\centering\arraybackslash}p{#1}}
\newcolumntype{M}[1]{>{\centering\arraybackslash}m{#1}}
\newcolumntype{R}[1]{>{\arraybackslash}m{#1}}

\definecolor{orange}{rgb}{1,0.5,0}
\definecolor{graynode}{RGB}{20,20,20}
\definecolor{crimsonred}{RGB}{220,20,60}
\definecolor{darkgraynode}{gray}{0.5}
\definecolor{lightgraynode}{gray}{0.8}

\usepackage{rotate}
\usepackage{capt-of}
\usepackage{setspace}
\usepackage{amssymb}
\usepackage{mathtools}
\usepackage{pifont} 
\newcommand{\cmark}{\ding{51}}
\newcommand{\xmark}{\ding{55}}

\definecolor{gray}{RGB}{20,20,20}
\definecolor{gray}{RGB}{0.7,0.7,0.7}
\definecolor{greencm}{RGB}{0,153,0}

\newcommand{\cm}{ {\color{greencm}\normalsize\cmark}}
\newcommand{\cmgray}{ {\color{gray}\normalsize\cmark}}
\newcommand{\xm}{ {\color{red}\normalsize\xmark}}

\usepackage{tabularx}
\usepackage{booktabs}

\usepackage{comment}
\usepackage{paralist}
\setcopyright{none}
\usepackage{nicefrac}
\definecolor{thelightblue}{RGB}{0,191,255}

\usepackage{nicefrac}

\usepackage{mathtools}
\usepackage{blkarray, bigstrut} 

\usepackage{tikz}
\usepackage{verbatim}
\usetikzlibrary{arrows}
\usetikzlibrary{shapes,snakes}
\usetikzlibrary{decorations.pathmorphing} 
\usetikzlibrary{fit}					
\usetikzlibrary{backgrounds}	

\makeatletter
\global\let\tikz@ensure@dollar@catcode=\relax
\makeatother

\usepackage{subfigure}

\definecolor{thelightblue}{RGB}{0,191,255}
\definecolor{theblue}{RGB}{0,0,180}

\usepackage{blkarray}
\usepackage{bigstrut, booktabs}

\makeatletter
\renewcommand*\env@matrix[1][*\c@MaxMatrixCols c]{
\hskip -\arraycolsep
\let\@ifnextchar\new@ifnextchar
\array{#1}}
\makeatother

\usepackage{booktabs} 
\usepackage{enumitem}

\usepackage{subfigure}
\usepackage{rotating}

\usepackage{multirow}

\definecolor{mydarkblue}{RGB}{0, 20, 159} 
\definecolor{mydarkblue}{rgb}{0,0.08,0.45} 
\usepackage{hyperref}
\hypersetup{colorlinks=true,urlcolor=mydarkblue, 
citecolor=mydarkblue,  linkcolor=mydarkblue}

\DeclareSymbolFont{cmbrightop}{OT1}{cmbr}{m}{n}
\DeclareMathSymbol{\sfPsi}{\mathalpha}{cmbrightop}{9}

\usepackage{tikz}
\usepackage{verbatim}
\usetikzlibrary{arrows}
\usetikzlibrary{shapes,snakes}
\usetikzlibrary{decorations.pathmorphing} 
\usetikzlibrary{fit}					
\usetikzlibrary{backgrounds}	

\definecolor{gray}{RGB}{150,150,150}
\definecolor{theblue}{RGB}{0, 20, 159} 

\definecolor{myyellow}{RGB}{255,255,204}
\definecolor{myred}{RGB}{255,204,204}
\definecolor{myblue}{RGB}{0,200,255}
\definecolor{mygreen}{RGB}{80,220,80}

\newcommand{\eg}{\emph{e.g.}}
\newcommand{\ie}{\emph{i.e.}}

\newcommand{\wrt}{\emph{w.r.t.}\ }

\newtheorem{Definition}{\hspace{-1em}\bfseries{Definition}}

\usepackage{balance}

\usepackage{epsfig}

\usepackage{graphicx}

\usepackage{amssymb}
\usepackage{amsmath}
\usepackage{amsfonts}

\usepackage{tabularx}
\usepackage{booktabs}
\usepackage{relsize}

\usepackage{numprint} 

\usepackage{setspace}
\graphicspath{{./}{./graphics/}}
\newcolumntype{H}{>{\setbox0=\hbox\bgroup}c<{\egroup}@{}}

\usepackage{algorithm}
\usepackage{algorithmicx}
\usepackage{algpseudocode}
\algblockdefx[parallel]{ParFor}{EndPar}[1][]{$\textbf{parallel for}$ #1 $\textbf{do}$}{$\textbf{end parallel}$}
\algrenewcommand{\alglinenumber}[1]{\fontsize{6.5}{7}\selectfont#1}
\algtext*{EndPar}

\algblockdefx[parallel]{parfor}{endpar}[1][]{$\textbf{parallel for}$ #1 $\textbf{do}$}{$\textbf{end parallel}$}
\algrenewcommand{\alglinenumber}[1]{\scriptsize#1:}

\algblockdefx[parallel]{parallelfor}{parallelend}
[1][]{\textbf{parallel for} #1}
{\textbf{end parallel}}

\usepackage{nicefrac}

\algnotext{EndFor}
\algnotext{EndIf}
\algnotext{EndProcedure}

\definecolor{dkgreen}{rgb}{0,0.6,0}
\definecolor{gray}{rgb}{0.5,0.5,0.5}
\definecolor{lightred}{rgb}{0.93,0.93,0.93}
\definecolor{lightred}{rgb}{0.83,0.83,0.83} 

\definecolor{lightgraydark}{rgb}{0.6,0.6,0.6}

\definecolor{lightblue}{rgb}{0.5,0.90,1.0}
\definecolor{lightgreen}{rgb}{0.5,0.92,0.5}
\definecolor{lightyellow}{rgb}{1,0.90,0.40}

\definecolor{verylightgreen}{RGB}	{204,255,204}
\definecolor{verylightred}{RGB}		{255,204,204}
\definecolor{verylightyellow}{RGB}		{255,255,204}

\definecolor{lightgreen}{RGB}	{204,255,204}
\definecolor{lightyellow}{RGB}		{255,255,204}

\definecolor{plotblue}{RGB}	{30,144,255}
\definecolor{plotgreen}{RGB}	{50,205,50}
\definecolor{plotred}{RGB}	{220,20,60}

\definecolor{myyellow}{RGB}{255,255,204}
\definecolor{myred}{RGB}{255,204,204}
\definecolor{lightblue}{RGB}{0,200,255}
\definecolor{mygreen}{RGB}{80,220,80}

\definecolor{gray}{RGB}{20,20,20}
\definecolor{greencm}{RGB}{0,153,0}

\definecolor{theblue}{RGB}{0,0,180}

\definecolor{matlabgreen}{rgb}{0,0.6,0}
\definecolor{matlabgray}{rgb}{0.5,0.5,0.5}
\definecolor{matlabmauve}{rgb}{0.58,0,0.82}

\usepackage[notheorems]{rr-math}

\begin{document}
\title{On Proximity and Structural Role-based Embeddings in Networks: Misconceptions, Techniques, and Applications}

\author{Ryan A. Rossi}
\orcid{1234-5678-9012-3456}
\affiliation{
\institution{Adobe Research}
\city{San Jose}
\state{CA}
\country{USA}
}
\email{rrossi@adobe.com}

\author{Di Jin}
\affiliation{
\institution{University of Michigan}
\city{Ann Arbor}
\state{MI}
\country{USA}
}
\email{dijin@umich.com}

\author{Sungchul Kim}
\affiliation{
\institution{Adobe Research}
\city{San Jose}
\state{CA}
\country{USA}
}
\email{sukim@adobe.com}

\author{Nesreen K. Ahmed}
\affiliation{
\institution{Intel Labs}
\city{Santa Clara}
\state{CA}
\country{USA}
}
\email{nesreen.k.ahmed@intel.com}

\author{Danai Koutra}
\affiliation{
\institution{University of Michigan}
\city{Ann Arbor}
\state{MI}
\country{USA}
}
\email{dkoutra@umich.com}

\author{John Boaz Lee}
\affiliation{
\institution{Worcester Polytechnic Institute}
\state{MA}
\country{USA}
}
\email{jtlee@wpi.edu}

\renewcommand{\shortauthors}{Rossi, Jin, Kim, Ahmed, Koutra, and Lee}

\begin{abstract}
Structural roles define sets of structurally similar nodes that are more similar to nodes inside the set than outside, whereas communities define sets of nodes with more connections inside the set than outside. 
Roles based on structural similarity and communities based on proximity are fundamentally different but important complementary notions.
Recently, the notion of structural roles has become increasingly important and has gained a lot of attention due to the proliferation of work on learning representations (node/edge embeddings) from graphs that preserve the notion of roles.
Unfortunately, recent work has sometimes confused the notion of structural roles and communities (based on proximity) leading to misleading or incorrect claims about the capabilities of network embedding methods.
As such, this paper seeks to clarify the misconceptions and key differences between structural roles and communities, and formalize the general mechanisms (\eg, random walks, feature diffusion) that give rise to community or role-based structural embeddings.
We theoretically prove that embedding methods based on these mechanisms result in either community or role-based structural embeddings.
These mechanisms are typically easy to identify and can help researchers quickly determine whether a method preserves community or role-based embeddings.
Furthermore, they also serve as a basis for developing new and improved methods for community or role-based structural embeddings.
Finally, we analyze and discuss applications and data characteristics where community or role-based embeddings are most appropriate.
\end{abstract}

\ccsdesc[500]{Computing methodologies~Artificial intelligence}
\ccsdesc[500]{Computing methodologies~Machine learning}
\ccsdesc[500]{Mathematics of computing~Graph algorithms}
\ccsdesc[500]{Mathematics of computing~Approximation algorithms}
\ccsdesc[300]{Mathematics of computing~Combinatorics}
\ccsdesc[300]{Mathematics of computing~Graph theory}
\ccsdesc[500]{Information systems~Data mining}
\ccsdesc[500]{Theory of computation~Graph algorithms analysis}
\ccsdesc[500]{Computing methodologies~Logical and relational learning}

\keywords{
Role-based structural embedding, roles, structural similarity, 
community-based embedding, 
proximity, 
role discovery, positions, 
structural embeddings,
structural node embeddings,
proximity-based node embeddings,
node embeddings,
network representation learning,
graphlets
}

\maketitle

\section{Introduction} \label{sec:intro}
Motivated by the proliferation of work on node representation learning and the confusion between the notions of communities and roles that existing methods capture, the goal of this manuscript is to clearly define and clarify the differences between community (proximity) and role-based (structural) embeddings.\footnote{Structural node embeddings and role-based embeddings are used synonymously in the literature. Similarly, proximity-based embeddings and community-based embeddings are also used synonymously.}
Towards this goal, we formalize these notions, discuss their differences, and show mathematically how various embedding mechanisms lead to either community or role-based structural embeddings. 
Given our definitions, we also categorize the existing embedding methods and discuss their suitability for a variety of downstream tasks.

In the following subsections, we begin by introducing the notion of communities and roles, which can be viewed as two different but complementary graph clustering problems.
Then, we discuss how these two fundamentally different notions give rise to community and role-based embeddings.
At the end of the introduction, we also present the scope of this article, its main contributions, and details about its organization.

\begin{figure}[t!]
\centering
\vspace{-3mm}
\hspace{-3mm}
\includegraphics[width=0.4\textwidth]{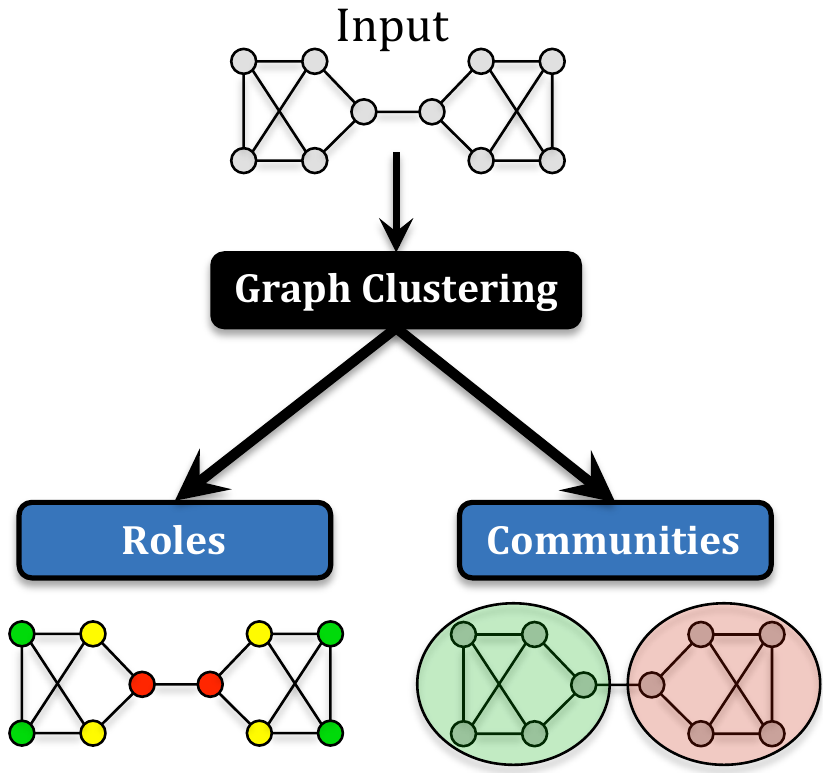}
\vspace{-0.2cm}
\caption{This taxonomy intuitively illustrates the fundamental differences between the notion of roles (which are based on structural similarity) and communities (based on density, cohesion, and small proximity/distance).
Table~\ref{table:roles-vs-comms} gives a summary of the key properties of communities and roles captured by embedding methods.
Note the input graph is the classical Borgatti-Everett network originally from~\cite{borgatti1992notions}. 
}
\label{fig:graph-clustering-taxonomy}
\end{figure}

\subsection{Communities and Roles} \label{sec:comm-and-roles-related-to-clustering}
Communities and roles lend themselves to many important real-world applications, which are discussed in the seminal survey on communities~\cite{fortunato2010community,graph-clustering-survey} and roles~\cite{roles2015-tkde}. 
They can be viewed as cases of general graph clustering, a problem that is fundamental to the analysis and understanding of graphs. 
Its main goal is to find a partition of nodes in an input graph.
We formalize the general definition of graph clustering as follows. 
\begin{Definition}[Graph Clustering] \label{def:graph-clustering}
A clustering $\mathcal{C} = \{C_1,\ldots,C_k\}$ of graph $G=(V,E)$ is a partition of the node set $V$ into non-empty subsets $C_i \subset V$ such that $V = \bigcup_{i}^{k} C_i$.
\end{Definition}
Definition~\ref{def:graph-clustering} does not specify the objective of the clustering, but simply that it is a partitioning of the vertex set $V$ into non-empty subsets $C_i$ such that $V = \bigcup_{i}^{k} C_i$.
Overall, there are two general objectives to graph clustering: (1) communities and (2) roles.  

\begin{Definition}[Communities] \label{def:communities}
Communities are sets of nodes with more connections inside the set than outside.\footnote{To capture the general notion of communities (Def.~\ref{def:communities}), there are many specific objective functions for the quality of communities~\cite{almeida2011there} including modularity~\cite{newman2004finding}, graph conductance~\cite{kannan2004clusterings}, among others~\cite{emmons2016analysis,almeida2011there}.}
That is, they are \emph{dense cohesive subsets} of vertices $\mathcal{C} = \{C_1,\ldots,C_k\}$.
A community $C_i \subseteq V$ is ``good" if the induced subgraph is dense (\ie, there are many edges between the vertices in $C_i$) and there are relatively few edges from $C_i$ to other vertices $\bar{C_i} = V \setminus C_i$~\cite{graph-clustering-survey}.
\end{Definition}

\begin{table*}[t!]
\renewcommand{\arraystretch}{1.2} 
\centering
\caption{
Roles and communities are fundamentally different but complementary notions.
Roles and communities are characterized below by their key properties.
}
\vspace{-3mm}
\label{table:roles-vs-comms}
\small
\footnotesize
\begin{center}
\begin{tabular*}{1.0\linewidth}{H p{0.53\linewidth} p{0.43\linewidth}}
\toprule 
& \textbf{Roles} 
& \textbf{Communities} \\
\midrule

\textbf{} & 
Roles form based on \emph{structural similarity}, \ie, nodes with similar structural properties (\eg, general subgraph patterns/graphlets)
&
Communities form based on node proximity/closeness (small distance), density, cohesiveness, sparse cuts 
\\
\midrule

\textbf{} & 
Roles are defined by structural properties/features
&
Communities are defined by node ids and proximity
\\ \midrule

& 
Roles generalize/transfer across networks (and can be used for graph-based transfer learning tasks) since they are defined by general structural properties/features 
& 
Communities do not generalize/transfer across networks (since based on node ids).
A community in $G$ has no meaning in another arbitrary graph $G^{\prime}$.
\\ \midrule

\textbf{} & 
Roles characterize nodes that are structurally similar with respect to their general connectivity and subgraph patterns (\eg, graphlets) and are independent of the distance/proximity to one another in the graph~\cite{roles2015-tkde}.
Hence, two nodes assigned to the same role can be in different communities or disconnected components of a single graph, or even different graphs 
&
Nodes in the same community should all be close to one another with small graph distance/proximity~\cite{fortunato2010community} 
\\

\bottomrule
\end{tabular*}
\end{center}
\end{table*}

Roles were first defined as classes of structurally equivalent nodes~\cite{lorrain1971structural}.
Intuitively, two nodes are {\emph structurally equivalent} if they are connected to the rest of the network in identical ways.
However, structural equivalence is far too strong and restrictive to be useful in practice.
Since then there have been many attempts to relax the criterion of equivalence, \eg, regular equivalence~\cite{white1983graph,everett1994regular}, stochastic equivalence \cite{holland1981exponential}.
For practical purposes, the notion of equivalence can be generally relaxed to get at some form of \emph{structural similarity}~\cite{roles2015-tkde}.
Roles may represent node (or edge)
connectivity patterns such as hub/star-center nodes, star-edge nodes, near-cliques or bridge nodes connecting different regions of the graph~\cite{ahmed2017roles,rolX}. 
More formally,

\begin{Definition}[Roles~\textnormal{\cite{roles2015-tkde}}] \label{def:roles}
Roles define sets of nodes that are more structurally similar to nodes inside the set than outside.
The term `structurally similar' refers to nodes that have similar structural properties, \eg, 
bridge-nodes (gatekeepers) that connect different communities.
The terms role and position are used interchangeably.\footnote{An equivalent definition of role is in terms of a role assignment function $r : V \rightarrow R$ that maps nodes to a set of roles $R$. The role assignment function $r$ induces a partition $\mathcal{C} = \{C_1,\ldots,C_k\}$ of $V$ by taking the inverse-images as sets/classes of nodes that play/have the same role.
Further, if $\sim$ is an equivalence relation (binary relation on $V$ that is reflexive, symmetric, and transitive), then the set of its equivalence classes is a partition of $V$ (and conversely).
Hence, it is equivalent to think of a role as a set of nodes (node partition), function (role assignment), or equivalence relation on $V$ since these are just different, but equivalent mathematical formulations for the concept of roles.}
\end{Definition}
In this work, the term \emph{structural similarity} (Definition~\ref{def:roles}) is reserved for the notion of roles as it implies nodes that have similar structural properties whereas the term proximity and density are reserved for communities.

Based on the definitions above, roles are complementary but fundamentally different to the notion of communities.
An intuitive example is shown in Figure~\ref{fig:graph-clustering-taxonomy} and their key differences are summarized in Table~\ref{table:roles-vs-comms}.
While communities capture cohesive/tightly-knit groups of nodes and nodes in the same community are close together (small graph distance or high proximity)~\cite{fortunato2010community}, roles characterize nodes that are structurally similar with respect to their general connectivity and subgraph patterns, and are independent of the distance/proximity to one another in the graph~\cite{roles2015-tkde}.
Hence, two nodes that share similar roles (\eg, star-center nodes) can be in different communities and even in two disconnected components of the graph.
Another difference is that communities are defined on a particular graph, whereas roles
capture a more general notion that represents structural patterns and are able to generalize across networks 
\ie, they can be learned on one network, and transferred to another,
whereas communities do not~\cite{roles2015-tkde,rolX}.
For instance, roles based on graphlets (or any structural features) can be naturally transferred to another graph as they are essentially general structural graph functions that can be computed on any arbitrary graph, independent of the specific nodes.
We refer the interested reader to Section~\ref{sec:role-based-embeddings} for further details and examples.

\subsection{Community- and Role-based Embeddings} 
Recently, there has been substantial work on learning node embeddings.\footnote{Embedding, features, and representation are considered synonymous and used interchangeably throughout this manuscript.}
Learning an appropriate feature-based representation of the graph (\eg, node/edge embeddings)
lies at the heart and success of many graph-based machine learning tasks.
In particular, they have proven to be important for many application tasks including
node and link classification~\cite{neville2000iterative,sen2008collective,mcdowell2009cautious,rossi2012drc,ember},
link prediction~\cite{deepGL,al2011survey,getoor:icdmw07,node2vec,deepwalk},
regression~\cite{rossi2014dynamical},
anomaly detection~\cite{akoglu2015graph,multilens}, 
dynamic network analysis~\cite{nicosia2013graph,kovanen2011temporal,nguyen2018continuous}, 
metric learning~\cite{ma-ahmed2018similarity,lee2020deep}, 
few-shot learning~\cite{garcia2018fewshot}, 
entity resolution/visitor stitching~\cite{node2bits,HONE,gilpin2013guided}, 
activity discovery~\cite{SafaviFSJWFKB20}, 
visualization and sensemaking~\cite{graphvis,pienta2015scalable,fang2017carina}, 
compression/graph summarization~\cite{role2vec,liu2018graph,multilens}, 
network alignment~\cite{koyuturk2006pairwise,heimann2018regal},
graph similarity~\cite{ma2019deep},
and graph classification~\cite{ying2018hierarchical,HeimannSK19,embed-graphsim,yan2019groupinn}.

While communities and roles are useful in themselves for many different and complementary applications, they have also become fundamentally important for learning embeddings that preserve the notion of community (proximity) or roles (structural similarity).
Indeed, many works claim to preserve the notion of communities~\cite{ComE,cavallari2019embedding,wang2017community}, roles~\cite{struc2vec,HONE,ember}, or even both~\cite{node2vec}.
The embedding/feature vectors given as output from an embedding method can be thought of as either community~\cite{fortunato2010community,henderson2010hcdf} or role membership vectors (assuming proper normalization)~\cite{Airoldi2008,rossi2013dbmm-wsdm,gilpin2013guided}. 
In this light, recent embedding methods can be seen as approaches for modeling communities \textit{or} (feature-based) roles~\cite{roles2015-tkde}.
For simplicity, we have described roles and communities in Section~\ref{sec:comm-and-roles-related-to-clustering} with respect to hard assignments.
However, more than a decade ago, methods for roles and communities that naturally output node embeddings have been investigated.
The node embeddings from these methods are sometimes referred to as node mixed-membership vectors.
In other words, community and role discovery are not different problems than node embeddings, since they both output node embeddings.

More recently, there has been an upsurge of interest in learning node embeddings 
that preserve structural roles~\cite{roles2015-tkde} (as opposed to communities based on proximity).
These works often claim to preserve structural equivalence~\cite{lorrain1971structural} or regular equivalence~\cite{white1983graph,sailer1978structural}. 
However, since the output of these methods are embeddings and not roles, these classical definitions are not appropriate since they are defined formally with respect to the graph as shown in Section~\ref{sec:roles}.
In particular, methods used to find role assignments that are regularly equivalent (or that preserve some other form of graph-based equivalence) use the graph directly and \textit{not} embeddings/features.
The recent work that learns embeddings is therefore more closely related to feature-based roles proposed by~\citet{roles2015-tkde}.
For instance, instead of leveraging the graph directly, feature-based roles are assigned based on (structural) feature representations (node embeddings) that appropriately describe the structural characteristics of the nodes in the graph.
Thus, the question becomes: given node embeddings learned from some method, do the embeddings preserve a feature/embedding-based role equivalence (or more generally, structural similarity) or do they preserve proximity (density, communities)?

Understanding whether a method preserves communities (proximity) or roles (structural similarity) helps identify and understand the applications and tasks where the embeddings might be useful.
For instance, if we know a method outputs embeddings that capture communities better, then we can already begin to understand the types of applications where such embeddings are likely to perform well, \eg, community-based embeddings work well for node classification on graphs with homophily (\ie, neighbors of a node are more likely to share the same label than not)~\cite{homophily-LaFond2010} whereas 
role-based embeddings work better for graphs with weak homophily or even heterophily.
We discuss applications of community and role-based embeddings in Section~\ref{sec:applications}.

\subsection{Scope of this Article} \label{sec:intro-scope}
This article formalizes the general mechanisms (techniques) that lie at the heart of nearly all existing embedding methods.
We show that these general mechanisms give rise to methods that learn community/proximity-based embeddings (Section~\ref{sec:node-edge-comm-embedding}) or role-based structural embeddings (Section~\ref{sec:role-based-embeddings}).
See Table~\ref{table:embedding-methods} for a summary of the key mechanisms behind community and structural role-based embeddings.
Moreover, we clarify the misconceptions surrounding these two different notions of embeddings.
This paper is not a survey of the abundance of work on communities~\cite{graph-clustering-survey,fortunato2010community} or roles~\cite{roles2015-tkde}, nor do we attempt to survey the abundance of work on graph embeddings/relational representation learning~\cite{rossi12jair,goyal2018graph,cai2018comprehensive,zhang2018network,wu2019comprehensive}.

\subsection{Main Contributions} \label{sec:intro-contributions}
The main contributions of this work are: 
\begin{itemize}
\item Formalizing the notion of communities and roles; and clarifying their fundamental differences

\item Proposing \emph{embedding-based node equivalences} for roles that are defined with respect to the learned node embeddings (feature vectors) as opposed to a graph $G$ as traditionally done.

\item Formalizing the general mechanisms that give rise to community or role-based structural embeddings. 
These mechanisms are typically easy to identify and can help researchers understand whether a method learns community or structural role-based embeddings.
Furthermore, they can also be used to develop new and better community or role-based embedding methods.

\item Theoretically demonstrating that embedding methods based on these mechanisms result in either community (proximity) or role-based structural embeddings.

\item Categorizing embedding methods into community or role-based by highlighting the general mechanism used by it and why it gives rise to such embeddings.

\item Analyzing and discussing the applications and data characteristics/assumptions where community-based or role-based embeddings are most appropriate.
\end{itemize}

\subsection{Organization of this Article} \label{sec:intro-organization}
The article is organized as follows: 
We first discuss background and preliminaries in Section~\ref{sec:prelim}.
In Section~\ref{sec:comm-and-roles-prelim}, we formalize the notion of communities and roles, discuss issues relating to false claims about these notions, and propose new feature/embedding-based equivalences for embedding methods.
In Sections~\ref{sec:node-edge-comm-embedding} and~\ref{sec:role-based-embeddings}, we formally describe the general mechanisms that lie at the heart of nearly all existing embedding methods.
We show that these general mechanisms give rise to methods that learn community/proximity-based embeddings (Section~\ref{sec:node-edge-comm-embedding}) or role-based structural embeddings (Section~\ref{sec:role-based-embeddings}).
For the general mechanisms behind community or role-based structural embeddings (summarized in Table~\ref{table:embedding-methods}), we theoretically demonstrate why they are community-based or role-based.
Section~\ref{sec:applications} discusses applications and the specific settings (\eg, data characteristics, problem setting/constraints)
that are well-suited for community or role-based embedding techniques.
Finally, Section~\ref{sec:conc} concludes.

\begin{table*}[t!]
\renewcommand{\arraystretch}{1.1} 
\small
\caption{General mechanisms that give rise to community (proximity) \emph{or} structural role-based embeddings.}
\vspace{-3mm}
\label{table:embedding-methods}
\small
\footnotesize
\begin{tabular*}{0.92\linewidth}{cl l H @{}H H HH H H@{}}
\toprule 
&&
& 
& 
&
& \textbf{Roles} 
& \textbf{Communities}
& 
\\

\multicolumn{1}{l}{\textbf{Embedding Type}}
& \textbf{General Mechanism}
& \textbf{Examples of Methods}
& 
&
& 
& 
& 
& 
\\
\midrule

\multirow{8}{*}{{\bfseries{\scshape{\fontsize{10}{12}\selectfont Community-based}}}} &
\multirow{6}{*}{\textbf{Random Walks} 
(Sec.~\ref{sec:node-edge-comm-embedding-walks})
} &
Spectral embedding~\cite{chung1997spectral} & 
\\

\multirow{8}{*}{(Section~\ref{sec:node-edge-comm-embedding})} & 
&
deepwalk~\cite{deepwalk} & 
\\

&&
node2vec~\cite{node2vec} & 
\\

&&
LINE~\cite{line} & 
\\

&&
GraRep~\cite{grarep} & 
\\

&&
ComE+~\cite{cavallari2019embedding} & 
\\
\cmidrule{2-7}

&
\multirow{3}{*}{\textbf{Feature Prop./Diffusion}
(Sec.~\ref{sec:node-edge-embedding-comm-feature-diffusion})} &
GCN~\cite{gcn} & 
\\

&&
GraphSage~\cite{graphsage} & 
\\

&&
MultiLENS~\cite{multilens} & \\
\midrule

\multirow{8}{*}{{\bfseries{\scshape{\fontsize{10}{12}\selectfont Role-based}}}} & 
\multirow{3}{*}{\textbf{Graphlets} 
(Sec.~\ref{sec:role-based-embeddings-graphlets})
} &
deepGL~\cite{deepGL} & 
\\

\multirow{8}{*}{
(Section~\ref{sec:role-based-embeddings})
} & 
&
MCN~\cite{lee18-higher-order-GCNs} & 
\\

&&
HONE~\cite{HONE} 
& \\
\cmidrule{2-6}

&
\multirow{3}{*}{\textbf{Feature-based Walks} 
(Sec.~\ref{sec:feature-based-walks})} &
role2vec~\cite{role2vec} 
& 
\\

&&
node2bits~\cite{node2bits}
&
\\

&&
SimSum~\cite{liu2019micro,liu2018semi}
& 
\\

\cmidrule{2-6}
&
\multirow{3}{*}{\textbf{Feature-based MF} 
(Sec.~\ref{sec:feature-based-matrix-factorization})} &
rolX~\cite{rolX}
& 
\\

&&
HERO~\cite{ahmed2017roles}
& \\

&&
EMBER~\cite{ember} & \\
\bottomrule
\end{tabular*}
\end{table*}

\section{Preliminaries} \label{sec:prelim}

Given a graph $G=(V, E)$ where $V$ represents the set of nodes and $E$ represents the set of edges, we define node and edge embedding as follows:

\begin{Definition}[Node Embedding] \label{def:node-embedding}
Node embedding aims to learn a function $f: V \rightarrow \mathbb{R}^{k}$ that maps each node to a $k$-dimensional embedding vector $\vx$ where $k \ll |V|$. 
\end{Definition}

\begin{Definition}[Edge Embedding] \label{def:edge-embedding}
Edge embedding aims to learn a function $f: E \rightarrow \mathbb{R}^{k}$ that maps an edge (node pair) to a $k$-dimensional embedding vector $\vx$ where $k \ll |E|$.
\end{Definition}
\noindent
Let $\mX$ denote the node (or edge) embedding/feature matrix where the rows represent nodes (or edges) and the columns represent (latent) features.
Hence, $\vx_i$ is the $k$-dimensional embedding/feature vector for the $i$-th node (edge). 

Many existing works derive edge embeddings based on the learned low-dimensional representations of nodes~\cite{deepwalk,node2vec} through element-wise operators such as \emph{average, Hadamard}, etc., so we categorize them together.
Approaches such as DeepGL~\cite{deepGL} that can learn edge embeddings directly from the graph are called \emph{direct edge embedding methods} and are discussed separately.

There is also a line of works that aim to learn an embedding vector for an entire graph:
\begin{Definition}[(Whole-) Graph Embedding] \label{def:graph-embedding}
Given a set of graphs, the goal is to learn a function $f : \mathcal{G} \rightarrow \RR^k$ that maps an entire input graph $G \in \mathcal{G}$ to a low-dimensional embedding vector $\vz$ of length $k$ where $\mathcal{G}$ is the input space of graphs.
Similar graphs (\eg, graphs belonging to the same class) should be embedded close to one another in the low $k$-dimensional space.
\end{Definition}\noindent
Some existing works in the literature aim to embed an induced subgraph
such as the subgraph rooted at a specific node~\cite{narayanan2016subgraph2vec,narayanan2017graph2vec}.
These methods can be easily applied to embed the entire graph by treating the input as a subset of the union of all graphs.
The graph embeddings can then be used as input for downstream applications such as 
graph classification~\cite{Lee18}, regression~\cite{Duvenaud15}, and anomaly detection~\cite{hu2016embedding}. 
We refer interested readers to the comprehensive review on the traditional graph embedding methods~\cite{fu2012graph}.

\section{Communities and Structural Roles} \label{sec:comm-and-roles-prelim}
This section formally defines the notions of communities and structural roles.
We then discuss and summarize the fundamental differences between these notions.
Despite the various applications and practical importance, the notion of \emph{structural roles} has only received a limited amount of attention~\cite{roles2015-tkde} compared to communities~\cite{graph-clustering-survey,Backstrom:2006,Chakrabarti:2006,newman2004fast,chen2010dense}.
As such, we discuss roles in significantly more detail and show that classical node equivalences for assigning roles with respect to $G$ are inappropriate for embeddings (by definition).

\subsection{Communities} \label{sec:communities}
While there are many different methods for finding communities (\eg, modularity maximization~\cite{newman2004finding}, cut-based methods~\cite{kannan2004clusterings}, among others~\cite{emmons2016analysis,almeida2011there}),
it is generally agreed that a subset of vertices $S \subseteq V$ is a ``good'' community if the induced subgraph is dense (\eg, many edges between the vertices in $S$) and there are relatively few edges from $S$ to the other vertices $\bar{S} = V \setminus S$~\cite{graph-clustering-survey}.
Let $E(S)$ denote the set of edges between vertices in $S$ (internal edges) and $E(S,\bar{S})$ be the set of edges between $S$ and $\bar{S}$ (cut set, that is, the set of edges that if removed would disconnect $S$ from $\bar{S}$).
Note $E(S,\bar{S})$ is the set of external edges, that is, the set of edges that have their origin in $S$ and destination in $\bar{S}$.
Clearly, $|E(S,\bar{S})|$ should be small relative to $|E(S)|$ and $E(\bar{S})$ for any ``good'' and reasonable community detection method.
An ideal situation is when communities are disjoint cliques.
A summary of the key properties of communities are as follows:
\begin{compactenum}
\item \textbf{Densely connected}: Nodes inside a community are more densely connected to nodes within the community than nodes in another community (By Definition~\ref{def:communities}).

\item \textbf{Proximity/closeness}: Nodes in the same community are close to one another in the graph in terms of distance/proximity. 

\item \textbf{Walks}:
Nodes in the same community have more walks to one another (\ie, ways of going from node $i$ to $j$) compared to nodes outside the community.
\end{compactenum}
Intuitively, both density and proximity are also closely related.
For instance, the more dense a community is in terms of the number of edges between nodes within the community, the more close (shorter distance/more walks) the nodes must be in the community, and vice-versa.
Both of these properties also imply more walks between nodes in the same community compared to nodes in another community.
See the seminal survey by~\citet{graph-clustering-survey} for discussion on other properties.
Note the term community-based embeddings and proximity-based embeddings are used interchangeably in the literature.

\subsection{Structural Roles} \label{sec:roles}
We first review and discuss the classical node equivalences used for structural roles.
These classical node equivalences (\eg, structural and regular equivalence~\cite{luczkovich2003defining}) are defined directly on the graph $G$ as opposed to the node embedding/feature matrix $\mX$.
Algorithms that return a structurally equivalent (or regularly equivalent) role assignment 
are known and have been widely studied~\cite{white1983graph,lorrain1971structural}.
However, since these node equivalences are defined with respect to $G$ and not embeddings, they cannot be used on embeddings/feature vectors, despite such claims in recent work.

In this work, we formalize the notion of \emph{structural similarity} and introduce node equivalences that \emph{can} be used on an embedding/feature representation of $G$.
These embedding-based node equivalences serve two main purposes.
First, they allow us to precisely formulate the notion of role mathematically which can be used to understand and theoretically analyze embedding methods and their representational power.
Second, algorithms based on these notions of feature-based node equivalences can be developed to find such roles.

We start by defining a role assignment, which is used later in the formalization of the different graph-based role equivalences (\eg, Definitions~\ref{def:structural-equiv}-\ref{def:regular-equiv}) and 
embedding-based role equivalences (Definitions~\ref{def:feature-based-equiv}-\ref{def:feature-based-struct-sim}) that lie at the heart of structural roles.
In particular, a role assignment $r$ of $V$ is a surjective mapping $r : V \rightarrow R$ onto a set $R$ of roles.
An assignment $r$ defines a partition\footnote{Recall a partition 
$\mathcal{C} = \{C_1, \ldots, C_k\}$ is a set of non-empty, disjoint subsets $C_i \subset V$ such that $V = \bigcup_{i=1}^{k} C_i$.}
$\mathcal{C} = \{C_1, \ldots, C_k\}$ of $V$ by taking the inverse-images as classes, \textit{i.e.}, $r^{-1}(s) = \{i \in V \,|\, r(i) = s\}, \forall s \in R$. 
Clearly, any role assignment $r$ is not useful alone, unless it satisfies one of the 
equivalences defined later in this section.
For instance, if $r$ satisfies regular equivalence (Definition~\ref{def:regular-equiv}), then we say the role assignment $r$ is regularly equivalent.

Given a node embedding/feature matrix $\mX$ from an arbitrary embedding/representation learning method $f$, we can find a role assignment $r : V \rightarrow R$ \emph{indirectly} using another method $\mathcal{M}$ (\eg, an approach that partitions nodes into disjoint sets $\mathcal{C} = \{C_1, \ldots, C_k\}$ such as $k$-means, or assigns nodes to classes).
However, since the role assignment $r$ is given by $\mathcal{M}$ and not $f$, it is difficult (if not impossible) to make any claims about $f$ with respect to classical notions of node equivalence such as structural (Definition~\ref{def:structural-equiv}) or regular equivalence (Definition~\ref{def:regular-equiv}).
In other words, there is no guarantee that a role assignment $r : V \rightarrow R$ is structurally or regularly equivalent since 
it is computed by some method $\mathcal{M}$ based on the learned node embedding/feature matrix $\mX$ only, despite the fact that the definitions of 
structural and regular equivalence involve the graph $G$ only (and more specifically, the neighbors of nodes) and not $\mX$.
Hence, the method $\mathcal{M}$ used to assign roles $r$ using $\mX$ would need to also consider the graph $G$ to guarantee that such a role assignment is structurally equivalent or regular equivalent; and it is unclear that $\mX$ would actually be useful in finding such a role assignment, since (efficient) algorithms that return such role assignments using only $G$ have been known for years~\cite{white1983graph,lorrain1971structural,borgatti2018analyzing,everett1991role,boyd1999relations}. 
In other words, there is no guarantee that the role assignments $r$ given by $\mathcal{M}$ based on node embeddings $\mX$ from 
$f$ will satisfy a graph-based equivalence such as structural equivalence or regular equivalence.
However, given a role assignment $r$ and a node equivalence (\eg, structural or regular equivalence), we can verify if the role assignment $r$ is a valid and proper assignment under the chosen node equivalence (\ie, the node equivalence holds for $r$).
Though, as mentioned above, this is not useful since there are methods to find such role assignments directly from the graph $G$, 
and in practice, such strict role assignments are typically not very useful.

\begin{Definition}[Role graph] \label{def:role-graph}
Let $G=(V,E)$ be a graph with role assignment $r : V \rightarrow R$, then $G_R=(R, E_R)$ is the role graph with vertex set $R$ (roles) and edge set $E_R \subseteq R \times R$ defined as:
\begin{equation} \label{eq:role-graph-edge-set}
E_R = \big\{\, (r_i, r_j) \,|\, (i,j) \in E \,\big\}
\end{equation}\noindent
where $G_R$ succinctly models the roles and the relationships between the roles $R$.
The role graph summarizes the essential structural properties of $G$ and thus can be viewed as a form of compression.
\end{Definition}

Role assignment definitions translate to partitions and equivalence relations for roles.
\begin{Definition}[Structural equivalence] \label{def:structural-equiv}
Let $G=(V,E)$ be a graph and $r : V \rightarrow R$ be a role assignment, then $r$ is \emph{strong structural} if equivalent vertices have the same neighbors. 
More formally, if $\forall i,j \in V$,
\begin{equation}
r_i=r_j \;\;\Longrightarrow\;\; N_i^{+}=N_j^{+} \,\wedge\, N_i^{-}=N_j^{-}
\end{equation}\noindent
where $N_i^{+}$ and $N_i^{-}$ are the out and in neighbors of $i$.
In other words, two nodes are structurally equivalent iff they are connected to the same neighbors.
Structural equivalence can only identify nodes close to each other in $G$. 
\end{Definition}
A few trivial examples of structurally equivalent nodes are star-edge nodes of a k-star graph, nodes in a k-clique graph, or nodes in a complete bipartite graph.
Structural equivalence is computationally and theoretically trivial. 
It is far too strict for general graphs and only nodes with distance at most two can be identified by it.
This has led to many slightly more useful relaxations including regular equivalence:

\begin{Definition}[Regular equivalence] \label{def:regular-equiv}
Let $r(W) = \{r(i) \,|\, i \in W\}$ be the \emph{role set} of $W \subseteq V$.
A role assignment $r : V \rightarrow R$ is \emph{regular} if $\forall i,j \in V$
\begin{equation}
r_i=r_j \;\;\Longrightarrow\;\; r(N_i^{+})=r(N_j^{+}) \,\wedge\, r(N_i^{-})=r(N_j^{-})
\end{equation}\noindent
where $r(N_i^{+})$ is the \emph{set of roles} from the neighbor set $N_i^{+}$.
Hence, $i$ and $j$ are \emph{regularly equivalent} iff they are connected to the same role equivalent neighbors (\ie, have the same set of roles from their neighbors).
\end{Definition}

\begin{Definition}[Exact role assignment] \label{def:exact-role-assignment}
A role assignment $r: V \rightarrow R$ is exact iff
\begin{equation}
r_i = r_j \,\;\Longrightarrow\;\, r(N_i) = r(N_j)
\end{equation}
where $N_i = \{j \in V \,|\, (i,j) \in E\}$ and $r(N_i)$ is the \emph{multi-set} of roles from set of neighbors $N_i$.
Hence, nodes of the same role must contain the same number of each of the other roles in their neighborhood.
\end{Definition}
Note the slight abuse of $r(N_i)$ in Definition~\ref{def:exact-role-assignment} to mean the \emph{multi-set} of roles from $N_i$ whereas in regular equivalence (Definition~\ref{def:regular-equiv}) and elsewhere it is simply the \emph{set} of roles.
An example of an exact role assignment is shown in Figure~\ref{fig:graph-clustering-taxonomy}.

\begin{Definition}[Strong structural role assignment] \label{def:exact-role-assignment}
Let $G=(V,E)$ and $G_R=(V_R,E_R)$ denote the role graph. 
A role assignment $r : V \rightarrow R$ is strong structural iff for all $i,j \in V$, there exists an edge $(r_i, r_j)$ in the role graph $G_R$ and $(i,j) \in E$.
\end{Definition}

All of the classical node equivalences are defined strictly using the graph $G$ alone, and not defined with respect to embeddings/features.
For instance, the formal definitions of structural and regular equivalence (Definition~\ref{def:structural-equiv}-\ref{def:regular-equiv}) involve only the sets of neighbors of nodes in $G$, and algorithms for finding such a role assignment $r$ that is structurally or regularly equivalent are known and require only the graph $G$ since that is all that is necessary by Definition~\ref{def:structural-equiv}-\ref{def:regular-equiv}.
For a summary of other classical graph-based node role equivalences that have the same issues, see~\cite{roles2015-tkde}.
Nevertheless, all such classical node equivalences that are formally defined \wrt the graph $G$ are obviously not helpful for assigning roles based on embeddings/features.
Furthermore, given node embeddings from any method, it is also not possible to use these classical graph-based node equivalences to make claims on whether the embeddings preserve the equivalence or not (by definition).

\medskip\smallskip\smallskip\smallskip
\noindent\textbf{From Equivalences on the Graph to Equivalences on Embeddings.}\;
While the classical node equivalences are defined \wrt the graph $G$ (and thus not useful for embeddings),
we now introduce the notion of an \emph{embedding-based node equivalence} defined \wrt node embeddings (as opposed to graph-based node equivalences discussed previously).
More importantly, we formalize the notion of structural similarity that serves as a basis for defining new node equivalences that involve node embeddings (features).
We begin by defining an equivalence relation for features:

\begin{Definition}[Feature-based/embedding equivalence] \label{def:feature-based-equiv}
A feature(embedding)-based equivalence is an equivalence relation $\sim$ between two structural feature/embedding vectors $\vx_i$ and $\vx_j$ for $i$ and $j$.
\end{Definition}

One of the most strict notions of node equivalence on embeddings/feature representation
is \emph{feature-based structural equivalence} defined as:
\begin{Definition}[Feature-based Structural Equivalence~\cite{roles2015-tkde}] \label{def:feature-based-struct-equiv}
Let $\mX$ be a \textbf{structural} embedding/feature matrix (\eg, features/embeddings that capture structural properties such as in/out/total degree, k-stars, k-paths, and other subgraph patterns/graphlets, etc).
A role assignment $r : V \rightarrow R$ is called \emph{feature-based structurally equivalent} if for all $v_i, v_j \in V$:
\begin{align}\label{eq:feature-based-struct-equiv}
r_i = r_j \;\; \Longrightarrow \;\;
\big[\forall k, 1 \leq k \leq d: x_{ik} = x_{jk}\big] 
\end{align}
\noindent
and $\mX$ are proper structural properties (and not based on proximity/cohesion)
where $x_{ik}$ is the $k$-\emph{th} feature value of node $i$.
Eq.~\ref{eq:feature-based-struct-equiv} is strict since two nodes belong to the same role iff they have identical feature vectors.
\end{Definition}
Notice a key difference between feature-based equivalences and the other graph-based equivalences is that there is no requirement on the \emph{neighbors} of a node.
This avoids roles being tied to one another based on proximity (cohesion/distance in the graph).
A more practical notion of roles is called \emph{feature-based structural similarity}~\cite{roles2015-tkde} 
and is a relaxation of the notion of feature-based structural equivalence (Definition~\ref{def:feature-based-struct-equiv}).
This notion replaces the strict requirement that nodes in the same roles have identical feature vectors by the requirement that nodes in the same roles must be \emph{$\epsilon$-structurally similar} for any $\epsilon$.
More formally, 

\begin{Definition}[Feature-based Structural Similarity]
\label{def:feature-based-struct-sim}
Let $\vx_i$ and $\vx_j$ be 
structural embeddings 
and $\mathbb{K}$ be a similarity function,
then we say node
$i$ and $j$ are \textbf{$\mathbf{\boldsymbol \epsilon}$-structurally similar} for any $\epsilon>0$ iff
\begin{compactenum}
\item $\vx_i$ and $\vx_j$ encode structural properties of $i$ and $j$ in $G$ (\eg, ``role-based'' features related to whether $i$ is on the periphery, star-edge, star-center/hub, bridge, near-clique, and so on)
\item $\vx_i$ and $\vx_j$ are not correlated with graph proximity and/or density (communities) in $G$
\item $\mathbb{K}(\vx_i,\vx_j) \geq 1-\epsilon$ for any $\epsilon>0$
\end{compactenum}
\end{Definition}
Intuitively, $i$ and $j$ are ``equivalent'' (which implies a partitioning/role assignment) iff $\mathbb{K}(\vx_i,\vx_j) \geq 1-\epsilon$ (\ie, they are $\epsilon$-structurally similar) and the features/embeddings are strictly structural and representative of the structural properties/topology in $G$ and not based on communities (\ie, proximity/closeness and density).

One can also add additional conditions to Definition~\ref{def:feature-based-struct-sim} to make it more strict (and possibly more useful for certain applications).
For instance, we can add an additional constraint that neighbors must not be of the same role.
While this constraint will strengthen condition 2 of Definition~\ref{def:feature-based-struct-sim}, it will also impact the roles we are able to capture, \eg, we would be unable to capture roles of nodes that represent near-cliques.
However, such a constraint will ensure that the features/embeddings do not simply capture communities, since if they did, then neighbors would likely be assigned to the same role.
We can relax this further by allowing one such role to have neighbors of the same role (\eg, to capture near-clique role).

\subsection{Discussion}
\label{subsec:disc}
In the context of embeddings, community-based methods embed nodes that are close in the graph (proximity, density) in a similar fashion whereas structural role-based methods embed nodes that are structurally similar (based on structural properties such as graphlets) such that \emph{structurally similar} nodes are close in some low $k$-dimensional space.
In some papers, there are misleading or false claims made about the resulting node embeddings.
For instance, some existing work claims that the resulting embeddings preserve the notion of structural equivalence or even regular equivalence.
Unfortunately, none of these notions are defined on the level of embeddings/features.
In fact, structural and regular equivalence are defined with respect to a graph $G$ and not node embeddings $\mX$ of the graph.
Thus it is impossible to apply such equivalences or make claims about such equivalences with respect to any arbitrary embeddings as shown and discussed in Section~\ref{sec:roles}.
As an aside, roles and communities can also be defined to allow a node or edge to belong to multiple roles and communities.
Typically, each node (edge) $i$ is assigned a weight $x_{ik}$ indicating the membership to the $k^{\rm th}$ cluster (community or role).
These are typically referred to as role or community membership models.
Overlapping communities is a special case of the above. 

In Table~\ref{table:qual-and-quant-comparison}, the general mechanisms behind community and role-based structural embeddings are summarized (which are discussed next in Section~\ref{sec:node-edge-comm-embedding}-\ref{sec:role-based-embeddings}) along with a few representative embedding methods from each of the mechanisms 
as well as the input and output of the community and role-based embedding methods. 
Recall from Section~\ref{sec:intro-scope} that this paper is not a survey of such embedding methods
(as this is outside the scope of this work), and thus, Table~\ref{table:qual-and-quant-comparison} shows only a few methods from each community and role-based mechanism. 
Moreover, the vast majority of embedding methods are community-based and thus, if we included all such methods, it would be highly skewed towards community-based embeddings.
As an aside, a recent work~\cite{ribeiro2019equivalence} claims to show the equivalence between what they call node embeddings and (structural) graph representations.\footnote{The term node embedding and representation are typically used interchangeably in the literature.}
However, the definitions introduced in that work 
deviate fundamentally from the definitions (and terminology) used in the literature and in this paper. 
Therefore, while the problems studied in~\cite{ribeiro2019equivalence} sound related, they are fundamentally different, and the corresponding findings and conclusions should not be confused.

\newcommand\TE{\rule{0pt}{2.0ex}}
\newcommand\BE{\rule[-1.1ex]{0pt}{0pt}}

{
\newcolumntype{C}{ >{\centering\arraybackslash} m{4cm} } 
\providecommand{\rotateDeg}{90}
\setlength{\tabcolsep}{1.2pt}
\providecommand{\rotDeg}{70}
\definecolor{verylightgreennew}{RGB}	{220,255,220}
\definecolor{verylightrednew}{RGB}		{255, 230, 230}
\definecolor{verylightreddarker}{HTML} {FFCBCB} 
\definecolor{verylightrednew}{RGB}		{255, 230, 230}
\definecolor{verylightrednewlighter}{RGB}		{255, 229, 239}
\definecolor{lightgraynew}{rgb}{0.9,0.9,0.9}
\definecolor{newgray}{RGB}{0.7,0.7,0.7}
\providecommand{\cellsz}{0.34cm} 
\providecommand{\cellszlg}{0.36cm} 
\renewcommand{\cm}{{\color{greencm}\normalsize\cmark}}
\renewcommand{\cmgray}{{\color{newgray}\normalsize\cmark}}
\renewcommand{\xm}{{\color{verylightreddarker}\normalsize\xmark}}
\newcommand\BBBBB{\rule[1.6ex]{0pt}{1.6ex}}
\newcommand\BBBBBB{\rule[-1.1ex]{0pt}{0pt}} 
\newcommand{\sysName}[1]{{\sf
\BBBBBB
#1
}}
\providecommand{\cellsomewhat}{
\BBBBB
\cmgray
\cellcolor{lightgraynew}
}
\providecommand{\cellno}{
\BBBBB
\xm
\cellcolor{verylightrednew}}
\providecommand{\cellyes}{
\BBBBB
\cm
\cellcolor{verylightgreennew}
}

\newcolumntype{H}{>{\setbox0=\hbox\bgroup}c<{\egroup}@{}} 
\begin{table}[t!]
\def\arraystretch{1.2} 
\scriptsize
\caption{
Qualitative and quantitative comparison of a few representative community and role-based graph embedding methods from each of the general mechanisms. 
}
\label{table:qual-and-quant-comparison}
\begin{minipage}{\columnwidth}
{\begin{center}
\vspace{-2mm}
\begin{tabular}
{
H
l
c 
!{\vrule width 0.8pt} 
P{\cellszlg} 
P{\cellszlg} 
H
!{\vrule width 0.6pt}
P{\cellsz} P{\cellsz} P{\cellsz}
P{\cellsz} P{\cellsz} P{\cellsz} 
!{\vrule width 0.6pt} 
P{\cellsz} P{\cellsz} 
P{\cellsz}  
H
!{\vrule width 0.6pt} 
P{\cellsz} P{\cellsz} 
P{\cellsz} 
P{\cellsz} 
P{\cellsz} 
!{\vrule width 0.6pt} 
H H H H
HHHH
!{\vrule width 0.8pt}
@{}
}

& & 
& \multicolumn{3}{c!{\vrule width 0.6pt}}{\textsc{}}
& \multicolumn{6}{c!{\vrule width 0.6pt}}{\textsc{Input}}
& \multicolumn{4}{c!{\vrule width 0.6pt}}{\textsc{Output}}
& \multicolumn{5}{c!{\vrule width 0.6pt}}{\textsc{Mechanism}}
\\

& \multicolumn{1}{l}{\textbf{Method}} & 
& 
\rotatebox{\rotateDeg}{\textbf{Community-based}} &
\rotatebox{\rotateDeg}{\textbf{Role-based}} &  
& 
\rotatebox{\rotateDeg}{\textbf{Homogeneous graph}} & 
\rotatebox{\rotateDeg}{\textbf{Heterogeneous graph}} &
\rotatebox{\rotateDeg}{\textbf{Temporal graph}} & 
\rotatebox{\rotateDeg}{\textbf{Weighted graph}} & 
\rotatebox{\rotateDeg}{\textbf{Directed graph}} & 
\rotatebox{\rotateDeg}{\textbf{Features/attributes}} & 
\rotatebox{\rotateDeg}{\textbf{Node embedding}} &  
\rotatebox{\rotateDeg}{\textbf{Direct edge embedding}} & 
\rotatebox{\rotateDeg}{\textbf{Graph embedding}} &
&

\rotatebox{\rotateDeg}{\textbf{Random walks}} &
\rotatebox{\rotateDeg}{\textbf{Feature Prop./Diffusion}} &
\rotatebox{\rotateDeg}{\textbf{Graphlet/Motif-based}} &
\rotatebox{\rotateDeg}{\textbf{Feature-based Walks}} &
\rotatebox{\rotateDeg}{\textbf{Feature-based MF}} &
&
&&&&
\\
\noalign{\hrule height 0.8pt}

& \sysName{$\mathsf{\sf \bf DeepWalk}$}~\cite{deepwalk}
&
& \cellyes 
& \cellno  
&   
& \cellyes 
& \cellno 
& \cellno 
& \cellyes
& \cellyes
& \cellno
& \cellyes 
& \cellno
& \cellno 
& \cellno  
& \cellyes   
& \cellno   
& \cellno  
& \cellno   
& \cellno   
& \cellno   
& \cellno  
& \cellno   
& \cellyes   
\\
\hline

& \sysName{$\mathsf{\sf \bf Node2vec}$}~\cite{node2vec}
&
& \cellyes 
& \cellno  
& \cellno  
& \cellyes 
& \cellno 
& \cellno 
& \cellyes 
& \cellyes 
& \cellno 
& \cellyes 
& \cellno  
& \cellno  
& \cellno  
& \cellyes   
& \cellno   
& \cellno  
& \cellno   
& \cellno   
& \cellno   
& \cellno  
& \cellno   
& \cellyes   
\\
\hline

& \sysName{$\mathsf{\sf \bf Metapath2vec}$}~\cite{dong2017metapath2vec}
&
& \cellyes 
& \cellno  
& \cellno  
& \cellyes 
& \cellyes 
& \cellno
& \cellyes 
& \cellno
& \cellno
& \cellyes 
& \cellno  
& \cellno  
& \cellno  
& \cellyes   
& \cellno   
& \cellno  
& \cellno   
& \cellno   
& \cellno   
& \cellno  
& \cellno   
& \cellyes   
\\
\hline

& \sysName{$\mathsf{\sf \bf CTDNE}$}~\cite{nguyen2018continuous}
&
& \cellyes 
& \cellno  
& \cellno  
& \cellyes 
& \cellno 
& \cellyes 
& \cellyes 
& \cellyes 
& \cellno
& \cellyes 
& \cellno  
& \cellno  
& \cellno  
& \cellyes   
& \cellno   
& \cellno  
& \cellno   
& \cellno   
& \cellno   
& \cellno  
& \cellno   
& \cellyes   
\\
\hline

& \sysName{$\mathsf{\sf \bf LINE}$}~\cite{line}
&
& \cellyes 
& \cellno  
& \cellno  
& \cellyes 
& \cellno 
& \cellno 
& \cellyes 
& \cellno 
& \cellno
& \cellyes 
& \cellno  
& \cellno  
& \cellno  
& \cellyes   
& \cellno   
& \cellno  
& \cellno   
& \cellno   
& \cellno   
& \cellno  
& \cellno   
& \cellyes   
\\
\hline

& \sysName{$\mathsf{\sf \bf S2S\text{-}AE}$}~\cite{taheri2018learning}
&
& \cellyes 
& \cellno  
& \cellno  
& \cellyes 
& \cellyes 
& \cellno 
& \cellyes 
& \cellno 
& \cellno 
& \cellno
& \cellno  
& \cellyes  
& \cellno  
& \cellyes   
& \cellno   
& \cellno  
& \cellno   
& \cellno   
& \cellno   
& \cellno  
& \cellno   
& \cellyes   
\\
\hline

& \sysName{$\mathsf{\sf \bf GraRep}$}~\cite{grarep}
&
& \cellyes 
& \cellno  
& \cellno  
& \cellyes 
& \cellno 
& \cellno 
& \cellyes 
& \cellno 
& \cellno
& \cellyes 
& \cellno  
& \cellno  
& \cellno  
& \cellyes   
& \cellno   
& \cellno  
& \cellno   
& \cellno   
& \cellno   
& \cellno  
& \cellno   
& \cellyes   
\\
\hline

& \sysName{$\mathsf{\sf \bf ComE+}$}~\cite{cavallari2019embedding}
&
& \cellyes 
& \cellno  
& \cellno  
& \cellyes 
& \cellno 
& \cellno 
& \cellyes 
& \cellno 
& \cellno
& \cellyes 
& \cellno  
& \cellno  
& \cellno  
& \cellyes   
& \cellno   
& \cellno  
& \cellno   
& \cellno   
& \cellno   
& \cellno  
& \cellno   
& \cellyes   
\\
\hline

& \sysName{$\mathsf{\sf \bf GCN}$}~\cite{gcn}
&
& \cellyes 
& \cellno  
& \cellno  
& \cellyes 
& \cellno 
& \cellno 
& \cellyes 
& \cellno
& \cellyes 
& \cellyes 
& \cellno  
& \cellno  
& \cellno  
& \cellno   
& \cellyes   
& \cellno  
& \cellno   
& \cellno   
& \cellno   
& \cellno  
& \cellno   
& \cellyes   
\\
\hline

& \sysName{$\mathsf{\sf \bf graphSAGE}$}~\cite{graphsage}
&
& \cellyes 
& \cellno  
& \cellno  
& \cellyes 
& \cellno 
& \cellno 
& \cellyes 
& \cellno
& \cellyes 
& \cellyes 
& \cellno  
& \cellno  
& \cellno  
& \cellno   
& \cellyes   
& \cellno  
& \cellno   
& \cellno   
& \cellyes   
& \cellno  
& \cellno   
& \cellyes   
\\
\hline

& \sysName{$\mathsf{\sf \bf MultiLENS}$}~\cite{multilens}
&
& \cellyes 
& \cellno  
& \cellno  
& \cellyes 
& \cellyes 
& \cellno 
& \cellyes 
& \cellyes
& \cellno
& \cellyes 
& \cellno  
& \cellno  
& \cellno  
& \cellno   
& \cellyes   
& \cellno  
& \cellno   
& \cellno   
& \cellyes   
& \cellyes  
& \cellyes   
& \cellyes   
\\
\hline

& \sysName{$\mathsf{\sf \bf DeepGL}$}~\cite{deepGL}
&
& \cellno 
& \cellyes  
& \cellno  
& \cellyes 
& \cellno 
& \cellno 
& \cellyes 
& \cellyes
& \cellyes 
& \cellyes 
& \cellyes  
& \cellno  
& \cellno  
& \cellno   
& \cellno   
& \cellyes  
& \cellno   
& \cellno   
& \cellyes   
& \cellyes  
& \cellyes   
& \cellyes   
\\
\hline

& \sysName{$\mathsf{\sf \bf MCN}$}~\cite{lee18-higher-order-GCNs}
&
& \cellno 
& \cellyes  
& \cellno  
& \cellyes 
& \cellno 
& \cellno 
& \cellyes 
& \cellyes
& \cellyes
& \cellyes 
& \cellno  
& \cellno  
& \cellno  
& \cellno   
& \cellno   
& \cellyes  
& \cellno   
& \cellno   
& \cellno   
& \cellno  
& \cellno   
& \cellyes   
\\
\hline

& \sysName{$\mathsf{\sf \bf HONE}$}~\cite{HONE}
&
& \cellno 
& \cellyes  
& \cellno  
& \cellyes 
& \cellyes 
& \cellno 
& \cellyes 
& \cellyes 
& \cellyes
& \cellyes 
& \cellno  
& \cellno  
& \cellno  
& \cellno   
& \cellno   
& \cellyes  
& \cellno   
& \cellno   
& \cellyes   
& \cellyes  
& \cellno   
& \cellyes   
\\
\hline

& \sysName{$\mathsf{\sf \bf role2vec}$}~\cite{role2vec}
&
& \cellno 
& \cellyes  
& \cellno  
& \cellyes 
& \cellno 
& \cellno 
& \cellyes
& \cellyes
& \cellyes 
& \cellyes 
& \cellno  
& \cellno  
& \cellno  
& \cellno   
& \cellno   
& \cellyes  
& \cellyes   
& \cellno   
& \cellyes   
& \cellyes  
& \cellno   
& \cellyes   
\\
\hline

& \sysName{$\mathsf{\sf \bf node2bits}$}~\cite{node2bits}
&
& \cellno 
& \cellyes  
& \cellno  
& \cellyes 
& \cellyes 
& \cellno 
& \cellyes 
& \cellyes
& \cellyes 
& \cellyes 
& \cellno  
& \cellno  
& \cellno  
& \cellno   
& \cellno   
& \cellno  
& \cellyes   
& \cellno   
& \cellno   
& \cellno  
& \cellyes   
& \cellyes   
\\
\hline

& \sysName{$\mathsf{\sf \bf rolX}$}~\cite{rolX}
&
& \cellno 
& \cellyes  
& \cellno  
& \cellyes 
& \cellno 
& \cellno 
& \cellyes 
& \cellyes
& \cellyes 
& \cellyes 
& \cellno  
& \cellno  
& \cellno  
& \cellno   
& \cellno   
& \cellno   
& \cellno   
& \cellyes  
& \cellno    
& \cellno    
& \cellyes   
& \cellyes   
\\
\hline

& \sysName{$\mathsf{\sf \bf  GLRD}$}~\cite{gilpin2013guided}
&
& \cellno 
& \cellyes  
& \cellno  
& \cellyes 
& \cellno 
& \cellno 
& \cellyes 
& \cellno  
& \cellyes 
& \cellyes 
& \cellno  
& \cellno  
& \cellno  
& \cellno   
& \cellno   
& \cellno   
& \cellno   
& \cellyes  
& \cellno    
& \cellno    
& \cellyes   
& \cellyes   
\\
\hline

& \sysName{$\mathsf{\sf \bf DBMM}$}~\cite{rossi2013dbmm-wsdm}
&
& \cellno 
& \cellyes  
& \cellno  
& \cellyes 
& \cellno  
& \cellyes 
& \cellyes 
& \cellno  
& \cellyes 
& \cellyes 
& \cellno  
& \cellno  
& \cellno  
& \cellno   
& \cellno   
& \cellno   
& \cellno   
& \cellyes  
& \cellno    
& \cellno    
& \cellyes   
& \cellyes   
\\
\hline

& \sysName{$\mathsf{\sf \bf struc2vec}$}~\cite{struc2vec}
&
& \cellno 
& \cellyes  
& \cellno  
& \cellyes 
& \cellno 
& \cellno 
& \cellno 
& \cellno 
& \cellno  
& \cellyes 
& \cellno  
& \cellno  
& \cellno  
& \cellno   
& \cellno   
& \cellno   
& \cellno   
& \cellyes  
& \cellno    
& \cellno    
& \cellno   
& \cellno   
\\
\hline

& \sysName{$\mathsf{\sf \bf xNetMF}$}~\cite{heimann2018regal}
&
& \cellno 
& \cellyes  
& \cellno  
& \cellyes 
& \cellno 
& \cellno 
& \cellno 
& \cellno
& \cellyes 
& \cellyes 
& \cellno  
& \cellno  
& \cellno  
& \cellno   
& \cellno   
& \cellno   
& \cellno   
& \cellyes  
& \cellno    
& \cellno    
& \cellyes   
& \cellyes   
\\
\hline

& \sysName{$\mathsf{\sf \bf EMBER}$}~\cite{ember}
&
& \cellno 
& \cellyes  
& \cellno  
& \cellyes 
& \cellno 
& \cellno 
& \cellyes 
& \cellyes
& \cellyes 
& \cellyes 
& \cellno  
& \cellno  
& \cellno  
& \cellno   
& \cellno   
& \cellno   
& \cellno   
& \cellyes  
& \cellno    
& \cellno    
& \cellyes   
& \cellyes   
\\
\hline

& \sysName{$\mathsf{\sf \bf HERO}$}~\cite{ahmed2017roles}
&
& \cellno 
& \cellyes  
& \cellno  
& \cellyes 
& \cellno  
& \cellyes 
& \cellyes 
& \cellno
& \cellyes 
& \cellyes 
& \cellyes  
& \cellno  
& \cellno  
& \cellno   
& \cellno   
& \cellyes   
& \cellno   
& \cellyes  
& \cellno    
& \cellno    
& \cellyes   
& \cellyes   
\\
\hline

\noalign{\hrule height 0.7pt}
\end{tabular}
\end{center}
}
\end{minipage}
\end{table}
}

\section{Community-based Embedding} \label{sec:node-edge-comm-embedding}
We discuss the two main general mechanisms behind existing \emph{community-based embedding methods}, namely,
random walks (Section~\ref{sec:node-edge-comm-embedding-walks}) and 
feature propagation/diffusion (Section~\ref{sec:node-edge-embedding-comm-feature-diffusion}).

\subsection{Random Walks} 
\label{sec:node-edge-comm-embedding-walks}
Random walks have been used as a basis in many community detection methods~\cite{van2000graph} for decades~\cite{Fortunato2010}.
Recent embedding approaches are based on traditional random walks and thus are \emph{unable} to capture roles (structural similarity) and instead capture the notion of communities~\cite{deepwalk,node2vec,ComE,cavallari2019embedding,dong2017metapath2vec} as shown later in this section.
Hence, these methods learn community/proximity-based embeddings, as opposed to structural role-based embeddings.
In particular, these methods embed nodes that are close to one another in the graph in a similar way and therefore are largely capturing the notion of communities as opposed to roles.
Recent empirical analysis shows that using random walks for embeddings primarily capture proximity among the vertices (see~\citet{goyal2018graph}), so that vertices that are close to one another in the graph (in terms of distance) are embedded together, \eg, vertices that belong to the same community are embedded similarly. 
In contrast, random walks will likely visit nearby vertices first, which makes them suitable for finding communities (based on proximity/density), rather than roles (structural similarity).
In fact, random walks are fundamental to many important community detection methods~\cite{van2000graph,graph-clustering-survey}.
Indeed, components of the eigenevector corresponding to the second eigenvalue of the transition matrix of a random walk on a graph provide proximity measures that indicate how long it takes for a walk to reach each vertex.
Obviously, vertices in the same community should be quickly reachable.
Furthermore, a random walk starting from a vertex in one community is more likely to remain in the community than to move to another community.
This is precisely the reason that random walks and communities are very closely related.

The normalized cut of a graph (used for community detection)
can be expressed in terms of the transition probabilities and the stationary distribution of a random walk in the graph~\cite{meila2001learning,meila2001random,chung1997spectral,dong2016sub,role2vec,orponen2008locally,orponen2005local}.
This formally links the mathematics of random walks to cut-based community detection methods.
Thus, random walks and communities are fundamentally tied.

The connection between walk-based embeddings and communities was formally shown by~\citet{role2vec}.
We summarize it below.
Suppose $\mathbf{P}$ is the transition matrix defined as
\begin{equation}
\mathbf{P}(u, v) = 
\left\{  
\begin{array}{lr}  
\frac{1}{d_u} & \text{if $u$ and $v$ are adjacent}\\  
0 & \text{otherwise.}
\end{array}  
\right.  
\end{equation}
The probability that a random walk $W(u)$ starting at $u$ visits a vertex $x$ at time $i$ is $\mathbf{e}_u\mathbf{P}^i\mathbf{e}_x^T$ where $\mathbf{e}_x$ is the unit vector having 1 in coordinate $x$ and $0$ in every other coordinate. For a directed edge $(u, v)$, the probability that a random walk $W(x)$ visits $u$ at time $i$ and then visits $v$ at time $i+1$ can be denoted as
\begin{equation*}
\frac{\mathbf{e}_x\mathbf{P}^i\mathbf{e}_u^T}{d_u}
\end{equation*}
Given an edge $(u, v)\in E$, let $I(u,v)$ denotes the total number of walks containing it.
With $d_v$ walks (each of which is of length $l$) starting at $v$, the sum of the probabilities that there exists a walk $W(x)$ that visits $(u,v)$ is
\begin{align}
I(u, v) &\le \sum^{l-1}_{i=0}\sum_{x}d_x\mathbf{e}_x\mathbf{P}^i\mathbf{e}_u^T/d_u\nonumber\\
&=\sum^{l-1}_{i=0}\mathbf{1D}\mP^i\ve^T_{u}/d_u\nonumber
=\sum^{l-1}_{i=0}\mathbf{1D}\ve^T_{u}/d_u\nonumber
=\sum^{l-1}_{i=0}1\nonumber
=l\nonumber
\end{align}
where $\mD$ is the degree matrix $\mD(u,u)=d_u$, $\mathbf{1}$ is the all-one vector and $\mathbf{1DP}^i = \mathbf{1D}$.
Therefore, if we start $d_u$ random walks from $u\in V$, the expectation of $I(u, v)$ is no more than $l$.

Let $C$ denote a community in the graph, $u, v \in C$ and $v'\in \bar{C}=V \setminus C$. 
The probability of the random walker staying in its community in the next step is:
\begin{equation}
\mathbf{P}(u, C) = \sum_{v\in C}\mathbf{P}(u, v) = \frac{d_{uC}}{d_u}
\end{equation}
where $d_{uC}$ denotes the number of edges originating from $u$ within the same community. Similarly, the probability of leaving the community is $\sum_{v'\in{\bar{C}}}\mathbf{P}(u, v')=\frac{d_{u\bar{C}}}{d_u}$.
By Definition~\ref{def:communities}, communities are densely connected with few edges across communities,
which implies that at time $i$ the probability of the random walker staying in the same community is larger than reaching nodes outside the community, \ie
\begin{equation}
d_{uC} > d_{u\bar{C}} \Rightarrow \mP(u, C) > \mP(u, \bar{C}) \hspace{0.4cm} \forall u \in C
\end{equation}
The above notion only reflects that at any time $i$, the random walk is more likely to sample neighbors of the same community. 
In order to measure the probability that all elements in a random walk stay in the same community, we introduce volume and conductance.
\begin{Definition}
Given a set of nodes $C \in V$ (partition of $V$), 
the volume of $C$ is $\mu(C)=\sum_{v\in C}d_v$. 
The conductance can then be computed as the ratio of its external edges over the minimum of $\mu(C)$ and $\mu(\bar{C})$:
\begin{equation}
\Phi(C) = \frac{|E(C, \bar{C})|}{\min(\mu(C), \mu(\bar{C}))}
\end{equation}\noindent
where $|E(C, \bar{C})|$ denotes the number of external edges between $C$ and $\bar{C} = V - C$
\end{Definition}
As shown by \citet{spielman2013local}, the probability that an $\ell$-step walk starting from a random vertex in $C$ stays entirely in $C$ is bounded by
\begin{equation}\label{eq:lower_1}
P_{C}^{\ell} \ge 1- \frac{\ell\Phi(C)}{2}
\end{equation}
This implies that if $\Phi(C)$ is small (which is usually the case for ``good commumities''), then the $\ell$-step walk will stay inside $C$ with fairly high probability. 
Further, while the probability to ``escape'' the community increases with longer walks, $\ell$ has to be comparable to $\frac{1}{\Phi(C)}$, which is generally a large value.
More recently, \citet{andersen2016almost} further shows that 
under certain mild assumptions,
this lower bound can be improved to $(1-\frac{\Phi(C)}{2})^\ell$. 
Nevertheless, both bounds indicate that random walks visit nodes in the same community with high probability.
Note that for disconnected communities their conductance values are always $0$, \ie, $\Phi(C) = 0$ since there are no external edges. 
Under this circumstance, both Equation~\eqref{eq:lower_1} and lower bound $(1-\frac{\Phi(C)}{2})^\ell$ produce probability $1$, which indicate that nodes from disconnected components can never be embedded in a similar fashion.

The above shows that random walks capture communities and thus any walk-based embedding method that uses either implicit walks (sequences of node ids) or explicit walks (number of walks between two nodes) outputs community-based embeddings.

\medskip\noindent\textbf{Explicit Walk-based Sampling}:
We first discuss methods that \emph{sample explicit walks} (sequences of node ids) from $G$ and then use these walks to derive embeddings.
A walk in $G$ is a sequence of nodes $v_1, v_2, \cdots, v_l$ s.t. $(v_{i}, v_{i+1}) \in E$, $\forall i$.
Note that the term walk-based sampling (or explicit walks)~\cite{ribeiro2012sampling,kolaczyk2014statistical} is used to distinguish techniques that sample explicit walks from the graph (representing sequences of node ids) from methods that are based on (implicit) walks, 
but do not explicitly sample them from the graph. 
The basic idea behind walk-based sampling is that nodes connecting with similar sets of neighbors (identified by ids) should be embedded closer. 
Therefore, these approaches first sample walks explicitly from the graph and then use these explicit sequences of ids to derive low-dimensional node embeddings that maximize the likelihood of predicting them.

Naturally, nodes and their neighbors in the walks are embedded closely in the vector space.
DeepWalk~\cite{deepwalk} is the first such method that leveraged explicit walks (sequences of node ids) to learn community-based embeddings.
Deepwalk employs Skip-Gram model to derive node embeddings that maximize the probability of neighbors identified in the explicit walks, \ie, 
\[
\displaystyle\arg \max_{f} Pr(v_{i-k},\cdots, v_{i-1}, v_{i+1}, \cdots,v_{i+k}|f(v_i)).
\]
As a result, nodes in the same community will be embedded closely by DeepWalk.
Based on DeepWalk, node2vec~\cite{node2vec} introduced a way to bias the random walks 
and claimed that both homophily (proximity/community-based embedding) and structural equivalency (Definition~\ref{def:structural-equiv}) can be preserved in the embeddings.
However, the notion of structural equivalence is defined only in terms of an actual role assignment, 
not an embedding, and therefore no claim can be made about whether an embedding is structurally equivalent (or regular equivalent).
Furthermore, as we showed above, random walks naturally give rise to community-based embeddings.
LINE~\cite{line} explicitly uses 1- and 2-hop neighbors (node contexts) to learn community/proximity-based node embeddings. 
LINE minimizes the KL-divergence between the first- and second-order joint probability distribution and the empirical distribution related to edge weights separately, and forms the output embeddings through concatenation. The embeddings derived by LINE incorporates local community information.
As indicated in~\cite{qiu2018network}, LINE can be seen as a special case of DeepWalk with the contextual size set to one.
More recently, ComE+~\cite{cavallari2019embedding} learns community-based embeddings by first sampling a fixed number of \emph{explicit paths} of a fixed length from every node, then leverages DeepWalk to obtain initial node embeddings.
Afterwards, the explicit paths are again used in an iterative fashion to obtain the final embeddings.

There are many extensions of DeepWalk to handle different types of graphs.
All such extensions also use explicit walks with node ids. 
One extension called metapath2vec~\cite{dong2017metapath2vec} is proposed to embed nodes in heterogeneous networks. This work relies on meta-path based random walk to capture contexts consisting of multiple node types following predefined meta-schemas. 
More recently, CTDNE~\cite{nguyen2018continuous} introduces the notion of temporal random walk and describes a general framework based on these temporal walks to learn temporally-valid embeddings at the finest temporal granularity. 
There are also walk-based sampling methods for graph embedding.
One such method by~\citet{taheri2018learning} generates multiple sequences including random walks, shortest paths between node pairs, and paths rooted at specific nodes to approximate the global graph structure and leverage sequence-to-sequence LSTM autoencoder to derive the embeddings. 
Other examples include Patchy-san~\cite{niepert2016learning} and random-walk-based sub2vec~\cite{adhikari2018sub2vec}. 

\medskip\noindent\textbf{Implicit Walk-based}:
Now we discuss implicit walk-based embeddings. 
These are characterized by the following property:
\begin{equation}\label{eq:k-step-matrix}
(\mA^k)_{ij} = \text{ number of walks of length k between } i \text{ and } j 
\end{equation}
Note that unlike embeddings that use \emph{explicit walks}, that is, sequences of node ids (\eg, DeepWalk, node2vec), implicit walk-based embeddings use the count of walks in some fashion.
Eq.~\ref{eq:k-step-matrix} obviously captures \emph{proximity} (\ie, communities) explicitly since $\mA^k$ is the number of walks of length $k$ between any two nodes.
Hence, the quantity itself describes the proximity between nodes.
Furthermore, $\mA^k_{ij}>0$ iff there is a walk from node $i$ to $j$ of length $k$, otherwise $(\mA^k)_{ij}=0$.
The above property is important, as this implies that nodes within the same community will have many such available walks compared to nodes between communities.
GraRep~\cite{grarep} is directly based on the above. 
In particular, GraRep computes $\mA^k$ for $k=1,\ldots,K$ and derives an embedding for each $\mA^k$ using SVD. The $K$ embeddings are then concatenated.

Besides $\mA^k$, we can also derive a matrix denoted as $\mA_k$ representing the sum of all walks up to length $k$ and use this for embedding.
More formally, the graph $G^k$ with the adjacency matrix denoted as $\mA_k$ given by the sum of the first $k$ powers of the adjacency matrix $\mA$ is:
\begin{align}\label{eq:sum-walks-up-to-length-k}
\mA_k & = \sum_{i=1}^k \mA^i \\
&= \mA + \mA^2 + \cdots + \mA^k \nonumber
\end{align}\noindent
where $(\mA_k)_{ij}>0$ iff there is a walk from $i$ to $j$ in at least $k$ steps and $(\mA_k)_{ij}=0$ if no such walk between $i$ and $j$ exists of length $1,\ldots,k$.
Setting $k=\mathsf{diam}(G)$ in Eq.~\ref{eq:sum-walks-up-to-length-k} gives a complete graph.

More generally, any matrix factorization method applied to $\mA$ \emph{directly} results in community-based embeddings as shown in~\cite{roles2015-tkde}.\footnote{This is the reason that feature-based role methods were proposed in~\cite{roles2015-tkde}, which avoid using $\mA$ directly, and instead derive a structural feature matrix $\mX$ that captures the structural properties (\eg, degree, triangles, betweenness, k-stars) of $G$ and then uses this matrix $\mX$ to derive roles.}
This is true for the adjacency matrix $\mA$ of $G$ or any matrix function of $\mA$ such as the normalized Laplacian $\mL$ or probability transition matrix $\mP$.
One such example is spectral embedding/clustering that computes the $k$ eigenvectors of the Laplacian matrix $\mL = \eye - \mD^{-1/2}\mA\mD^{-1/2}$ of $G$~\cite{spectral,ng2002spectral}.
The intuition is the same as above.
Another example is HOPE~\cite{ou2016asymmetric}, which proposes 4 different ways to measure the proximity, which are Katz Index, personalized PageRank, Common neighbors and Adamic-Adar.
TADW~\cite{yang2015network}, HSCA~\cite{zhang2016homophily} leverage Pointwise Mutal Information (PMI) of word-context pair to denote proximity.
CMF~\cite{zhao2015representation} leverages Positive Pointwise Mutal information (PPMI) by omitting unrelated pairs of nodes with negative PMI values.
Modularized NMF (M-NMF)~\cite{wang2017community} is an implicit walk-based approach for deriving community/proximity-based embeddings by jointly optimizing an NMF-based model with a modularity-based community detection model.
All of these methods are community-based.

These embeddings have connections to eigenvectors and in particular the principle eigenvector.
The proof of the Ergodic theorem is most frequently given as an application of the Perron Frobenius theorem, which states that the probabilities of being at a node are given as the coefficients of the \emph{principal eigenvector} of the stochastic transition matrix $\mP$ associated to the Markov chain computed as follows:
\begin{equation}
\lim_{k\rightarrow\infty} \mP^k\ve
\end{equation}
\noindent 
where $\ve$ is the unit vector.
Recall that $\mA\ve$ gives the degree vector whereas $\mA^2\ve$ is the number of walks of length 2, and so on.
In general, the operation $\mA\ve$ is essentially equivalent to a single BFS 
iteration (over all nodes), see~\cite{kepner2011graph} for more details.\footnote{
BFS and DFS are general graph traversal/search techniques used in the implementation of graph algorithms.
These techniques are simply different ways to visit nodes in the graph, and
are often used in the implementation of \emph{both} community- and structural role-based embeddings.}
The above has been used for decades in a variety of seminal community detection and community-based embedding methods~\cite{gibson1998inferring,graph-clustering-survey,andersen2006local}.

\subsection{Feature Propagation/Diffusion} \label{sec:node-edge-embedding-comm-feature-diffusion}
While most community-based embeddings arise from explicit or implicit walks (Section~\ref{sec:node-edge-comm-embedding-walks}), there are also many methods that use feature diffusions (\ie, feature propagation) to learn community-based embeddings.
These methods are fundamentally tied to communities as they are still related to walk-based methods (which we formally showed in Section~\ref{sec:node-edge-comm-embedding-walks} that such walk-based methods are fundamentally community-based). 
The only difference is that features are diffused through the neighborhoods.
Thus, as $k \rightarrow \infty$, for any $(\mA\mX)^k$, then the features are smoothed over the graph.
Using Eq.~\ref{eq:lower_1}, we can see that nodes within the same community will have similar embeddings since the diffusion and resulting features primarily stay within the same community by definition.
Thus, the nodes within the same community become more and more similar as features are diffused from further away (but primarily from nodes within the same community), making all such nodes in the same community appear increasingly similar to one another.
This is precisely why such graph diffusion lies at the heart of many community detection methods such as heat kernel~\cite{kloster2014heat}, PageRank communities~\cite{andersen2006local}, among many others~\cite{kondor2002diffusion,raghavan2007near}.
Furthermore, these methods typically rely on selecting a good seed set of nodes to begin such diffusions.
In theory, a good seed set should contain one or more vertices from each ``community'', otherwise, some communities will be missed for precisely the same reason (it is unlikely that a walk starting from one community will end up in another community) as shown above in Eq.~\ref{eq:lower_1}.

In general, propagation/diffusion-based methods make the assumption that some initial attributes are given as input and stored in $\mX$.
For instance, we may associate with each node in a social network features that were taken from the corresponding user's profile. 
Node embeddings are then generated via a $k$-step diffusion process. 
At each step, a node's features are diffused to its immediate neighbors and, after $k$ rounds, each node obtains an embedding which is essentially an aggregation of the information in its $k$-order neighborhood.
While the diffusion process is dependent on walks between node pairs, methods falling into this category do not explicitly leverage walks to approximate the graph structure. 
Instead, they propagate information over the graph structure up to $k$-orders, and characterize individual nodes by collecting the diffused feature values in the neighborhood~\cite{deepGL,xiang2018feature}.
Thus, embeddings based on feature diffusion 
are community-based as the diffusion process is fundamentally tied to proximity in the graph as opposed to structural properties of nodes.
See Table~\ref{table:qual-and-quant-comparison} for a summary of a few representative network embedding methods based on feature propagation/diffusion.

The general form of this process is denoted as
\begin{equation}
\tilde{\mathbf{X}} = \Psi(\mA,\mX)
\label{eq:general-diffusion}
\end{equation}
where $\Psi$ denotes the feature expansion or diffusion function and $\tilde{\mX}$ denotes the expanded features. 
In the simplest case, a feature matrix propagating over the graph structure can be written as the standard form of Laplacian smoothing:
\begin{equation}
\tilde{\mX}^{(t)} = \mD^{-1}\mathbf{A}\tilde{\mX}^{(t-1)} 
= (\mD^{-1}\mA)^t\tilde{\mX}^{(0)} = (\mD^{-1}\mA)^t\mX
\label{eq_simple}
\end{equation}
where $\mathbf{D}$ is the diagonal degree matrix and $t$ represents iteration $t$ of the diffusion process. $\tilde{\mathbf{X}}^{(0)}$ is generally set to be the initial feature matrix $\mathbf{X}\in\mathbb{R}^{N\times F}$, \ie, $\tilde{\mathbf{X}}^{(0)} = \mathbf{X}$.

More complex feature diffusion processes can be denoted through the Laplacian diffusion process:
\begin{equation}
\tilde{\mathbf{X}}^{(t)} = (1-\theta)\mathbf{L}\tilde{\mathbf{X}}^{(t-1)} + \theta\mathbf{X}
\label{eq_complex}
\end{equation}
where $\theta$ controls the weighting between features of a node itself and its neighbors. $\mathbf{L}$ is the normalized Laplacian:
\begin{equation}
\label{eq:lap}
\mathbf{L} = \eye - \mathbf{D}^{-\frac{1}{2}}\mathbf{A}\mathbf{D}^{-\frac{1}{2}}
\end{equation}
\noindent
The Laplacian smoothing process generates new features as the weighted average given a specific node itself and its neighbors. 

Many of the recent propagation or diffusion-based methods~\cite{graphsage,gcn,gat,kipf2016variational} also incorporate trainable parameters into the diffusion process 
to learn better community/proximity-based embeddings. 
For instance, the step-wise diffusion for GCN can be defined as follows:
\begin{equation}
\tilde{\mathbf{X}}^{(t)} = \sigma \left( \mathbf{D}^{-\frac{1}{2}} \mathbf{A} \mathbf{D}^{-\frac{1}{2}} \tilde{\mathbf{X}}^{(t-1)} \mathbf{W}^{(t)} \right)
\label{eq_simple}
\end{equation}
\noindent where $\mathbf{W}^{(t)}$ is the weight matrix for step $t$ and $\sigma$ is a non-linearity. 
Depending on specific propagation configuration, there is a number of recent works that fall under this category. 
For example, GAE (and its variant using variational training manner, VGAE)~\cite{kipf2016variational} leverages GCN to encode node features, and then uses a simple decoder to reconstruct the graph adjacency matrix so that the loss can be minimized. 
DySAT~\cite{you2019g2sat} casts the Boolean Satisfiability (SAT) problem as a problem of deriving latent bipartite graph representations and provides a solution in the GCN feature aggregation manner.
HGCN~\cite{chami2019hyperbolic} extends GCN into the hyperbolic space so that node features are learned with less distortion.
Some other variants are devoted to capturing both spatial and temporal dependency between nodes in
the graph, such as ~\cite{wu2019graph}.
In order to gain interpretability of these models, GNNExplainer~\cite{ying2019gnnexplainer} was proposed as a model-agnostic approach to explain the prediction/inference on machine learning tasks, while GroupINN~\cite{yan2019groupinn} introduced a grouping or summarization layer that explains the classification by exposing the most relevant edges.
PRUNE~\cite{prune} is another recent proximity/community-based embedding method that uses a Siamese neural network structure to preserve proximity among the nodes.

Now we show that repeated application of a smoothing operator results in the features converging to the same quantity.
In other words, the features of nodes within the same community become indistinguishable from one another as the number of iterations (feature propagations) become large.
In particular, we prove that by iteratively applying the smoothing operator $\mD^{-1}\mA$, the feature vectors associated with nodes in a connected graph $G$ will converge in the end. 
The same applies when $\mD^{-\frac{1}{2}}\mA\mD^{-\frac{1}{2}}$ is used as a smoothing operator over every node and its neighbors in the graph.

\begin{theorem}\label{thm:feature-diffusion-conv}
Assuming $G$ is connected and non-bipartite, then for any feature/embedding matrix $\mX \in \RR^{n\times F}$:
\begin{equation}\label{eq:diffusion-Lrw}
\lim_{t\rightarrow\infty}(\mD^{-1}\mA)^{t}\mX
= \mathbf{1}\mathbf{y}^T
\end{equation}
and
\begin{equation}\label{eq:diffusion-Lnorm}
\lim_{t\rightarrow\infty} (\mD^{-\frac{1}{2}}\mA\mD^{-\frac{1}{2}})^{t}\mX = \mD^{-\frac{1}{2}}\mathbf{1}\mathbf{y}^T
\end{equation}
where $\mathbf{y}\in\mathbb{R}^{F}$.
Hence, Eq.~\ref{eq:diffusion-Lrw} converges to identical feature vectors for all nodes whereas the features of nodes smoothed using Eq.~\ref{eq:diffusion-Lnorm} (normalized Laplacian) converge to be proportional to the square root of the node degree.
\end{theorem}
Note that $\mD^{-1}\mA$ in Eq.~\ref{eq:diffusion-Lrw} is for the random walk Laplacian matrix whereas $\mD^{-\frac{1}{2}}\mA\mD^{-\frac{1}{2}}$ in Eq.~\ref{eq:diffusion-Lnorm} is for the normalized Laplacian.

\begin{proof}
Let $\mL_{rw} =\mI - \mD^{-1}\mA$ and $\mL = \mI - \mD^{-\frac{1}{2}}\mA\mD^{-\frac{1}{2}}$ (Eq.~\eqref{eq:lap}).
$\mL_{rw}$ and $\mL$ have the same $n$ eigenvalues by multiplicity with different eigenvectors with eigenvalues in $[0,2)$~\cite{chung1997spectral}.
The eigenspaces corresponding to eigenvalue 0 are thus spanned by $\mathbf{1}$ and $\mD^{-\frac{1}{2}}\mathbf{1}$, respectively.
Thus the eigenvalues of $\mD^{-1}\mA$ ($\mI - \mL_{rw}$) and $ \mD^{-\frac{1}{2}}\mA\mD^{-\frac{1}{2}}$ ($\mI - \mL$) would fall into $(-1, 1]$. 
Since the absolute value of the eigenvalues are less than or equal to $1$, the repeating multiplication will converge to the largest eigenvector corresponding to eigenvalue $1$, which is $\mathbf{1}$ and $\mD^{-\frac{1}{2}}\mathbf{1}$, respectively.
\end{proof}

The above obviously holds for $k = |\mathcal{C}|$ communities $\mathcal{C}=\{C_1,\ldots,C_k\}$ such that 
$|E(C_i,C_j)|=0, \forall i,j$, \ie, there are no edges between any pair of the communities.
This is an extreme case.
However, when the number of feature propagations is small, it is easy to see that the resulting embedding vectors of nodes within the same community become more and more similar due to the smoothing of the nodes within the same community.
Intuitively, 
when $t$ is small, then after $t$ feature propagations, the resulting diffused feature vectors of nodes within the same community are more similar to each other than to the diffused feature vectors of nodes in another community.
This occurs since the nodes inside a community are not as impacted by the features in another community due to the sparse edges between communities, \ie, $|E(C_i,C_j)| \leq |E(C_i)|,|E(C_j)|$, hence the number of edges between nodes in the same community is significantly larger than the number of edges between communities (by Definition~\ref{def:communities}).
Thus, nodes in the same community become more and more similar to each other.

\begin{figure}[t!]
\centering
\hspace{-5mm}
\subfigure[Initial random feature]{
\includegraphics[width=0.25\linewidth]{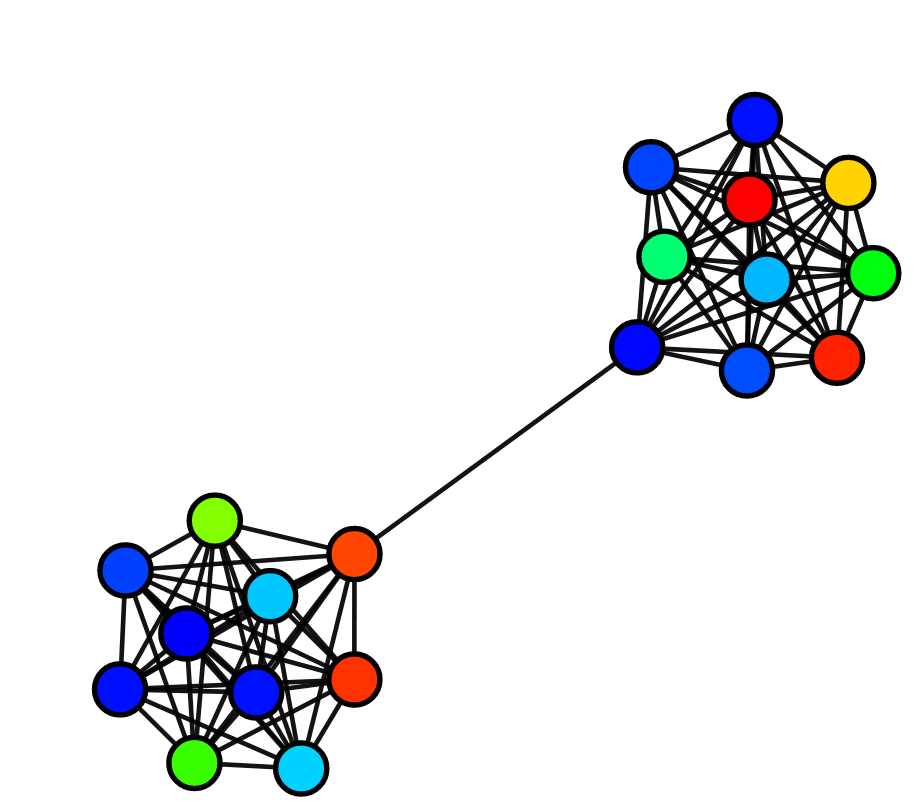} 
\label{fig:feature-diffusion-example-X}
}
~~
\subfigure[After 1 feature diffusion]{
\includegraphics[width=0.25\linewidth]{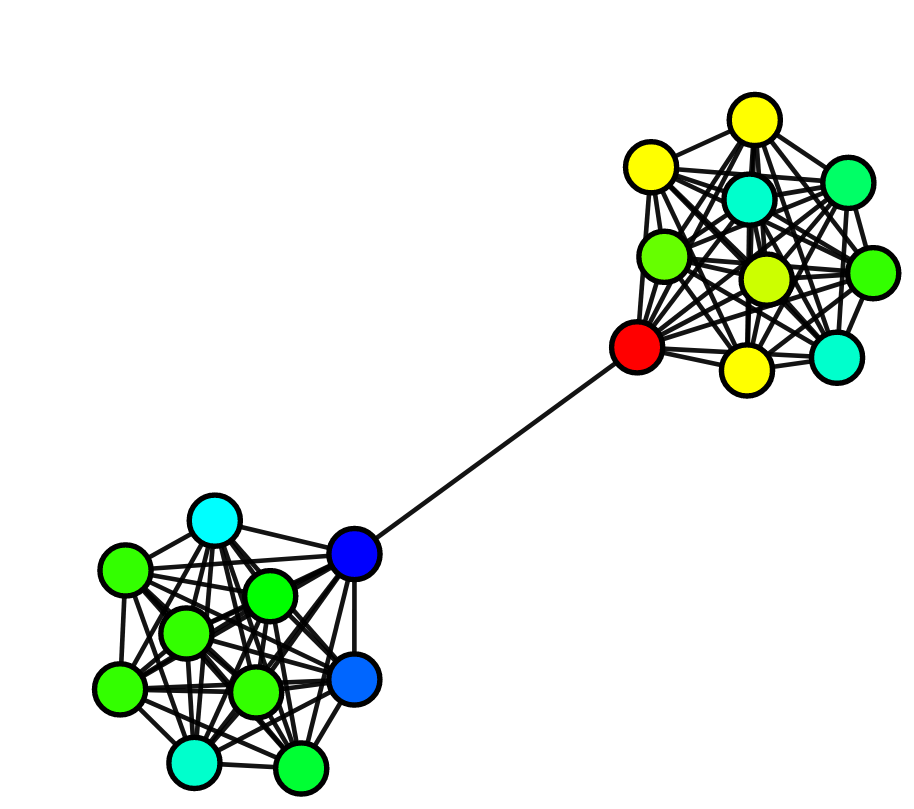} 
\label{fig:feature-diffusion-example-X1}
}
~~
\subfigure[After 3 feature diffusions]{
\includegraphics[width=0.25\linewidth]{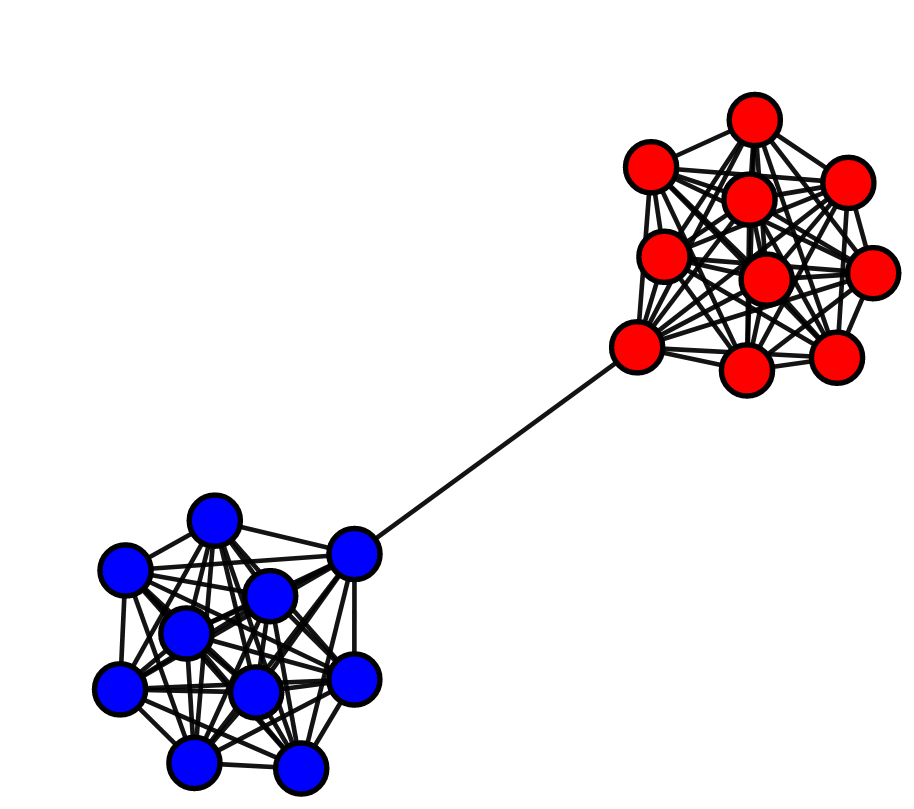} 
\label{fig:feature-diffusion-example-X3}
}

\subfigure[Initial random feature]{
\includegraphics[width=0.25\linewidth]{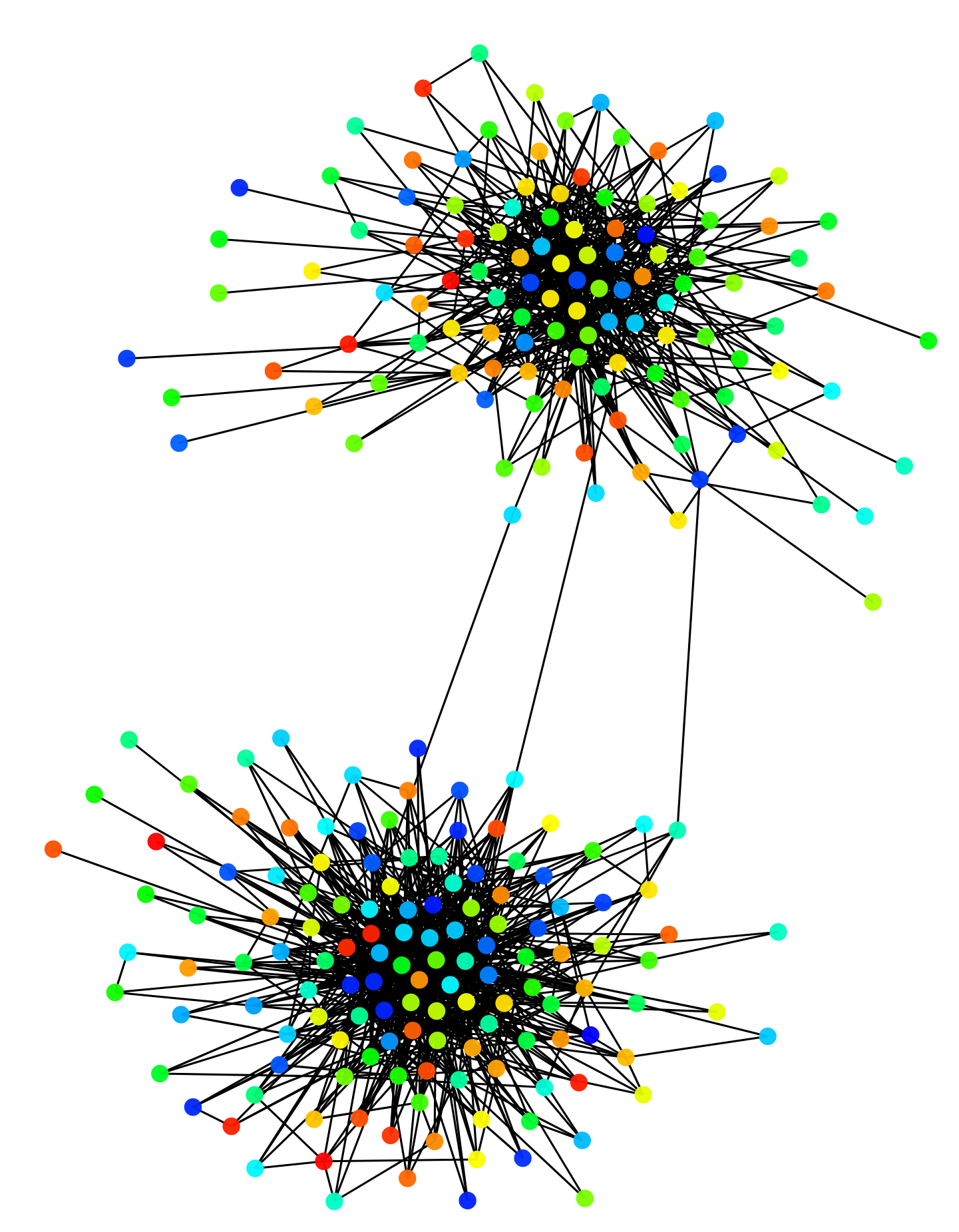} 
\label{fig:feature-diffusion-example-X}
}
~~
\subfigure[After 3 feature diffusions]{
\includegraphics[width=0.25\linewidth]{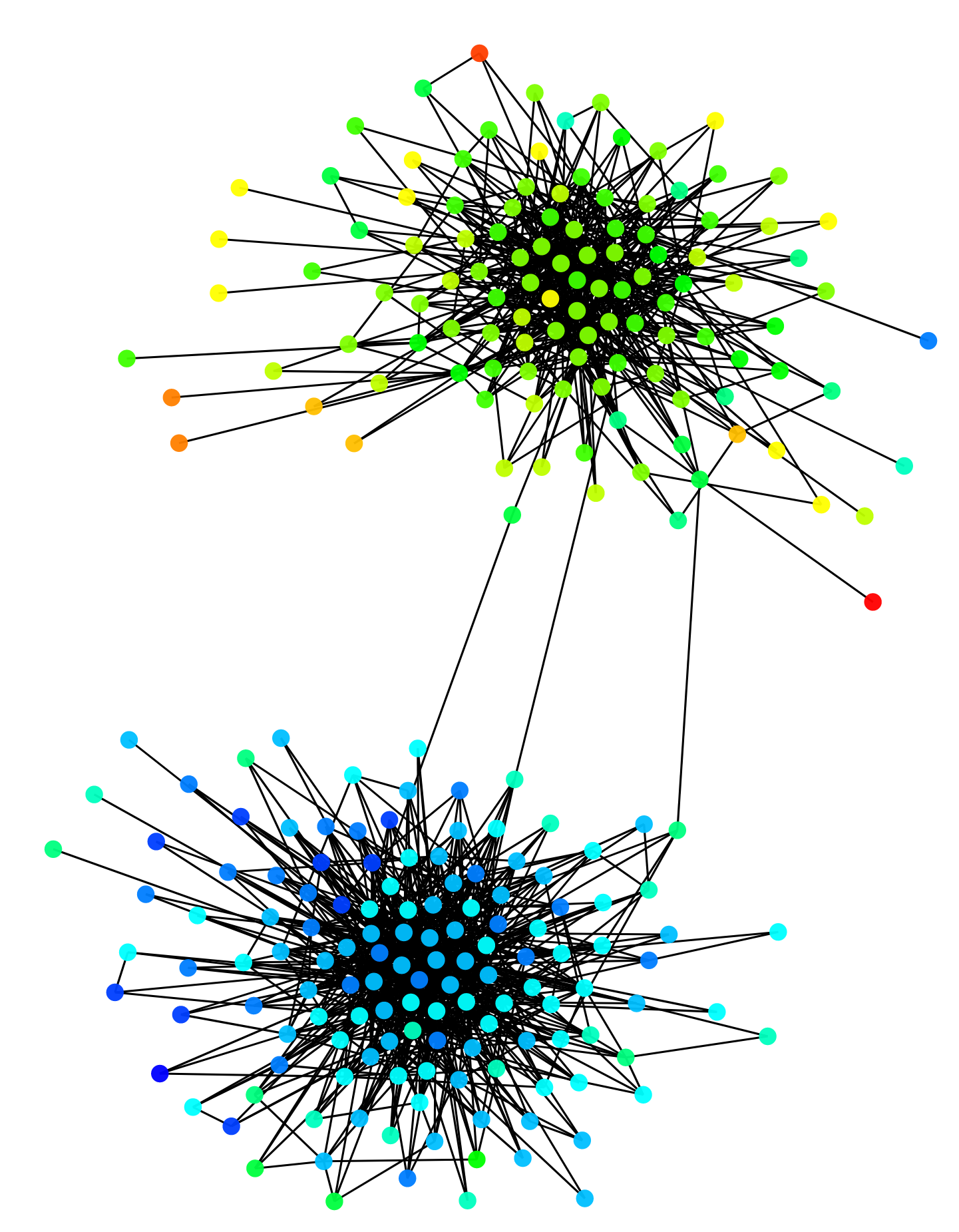} 
\label{fig:feature-diffusion-example-X3}
}
~~
\subfigure[After 5 feature diffusions]{
\includegraphics[width=0.25\linewidth]{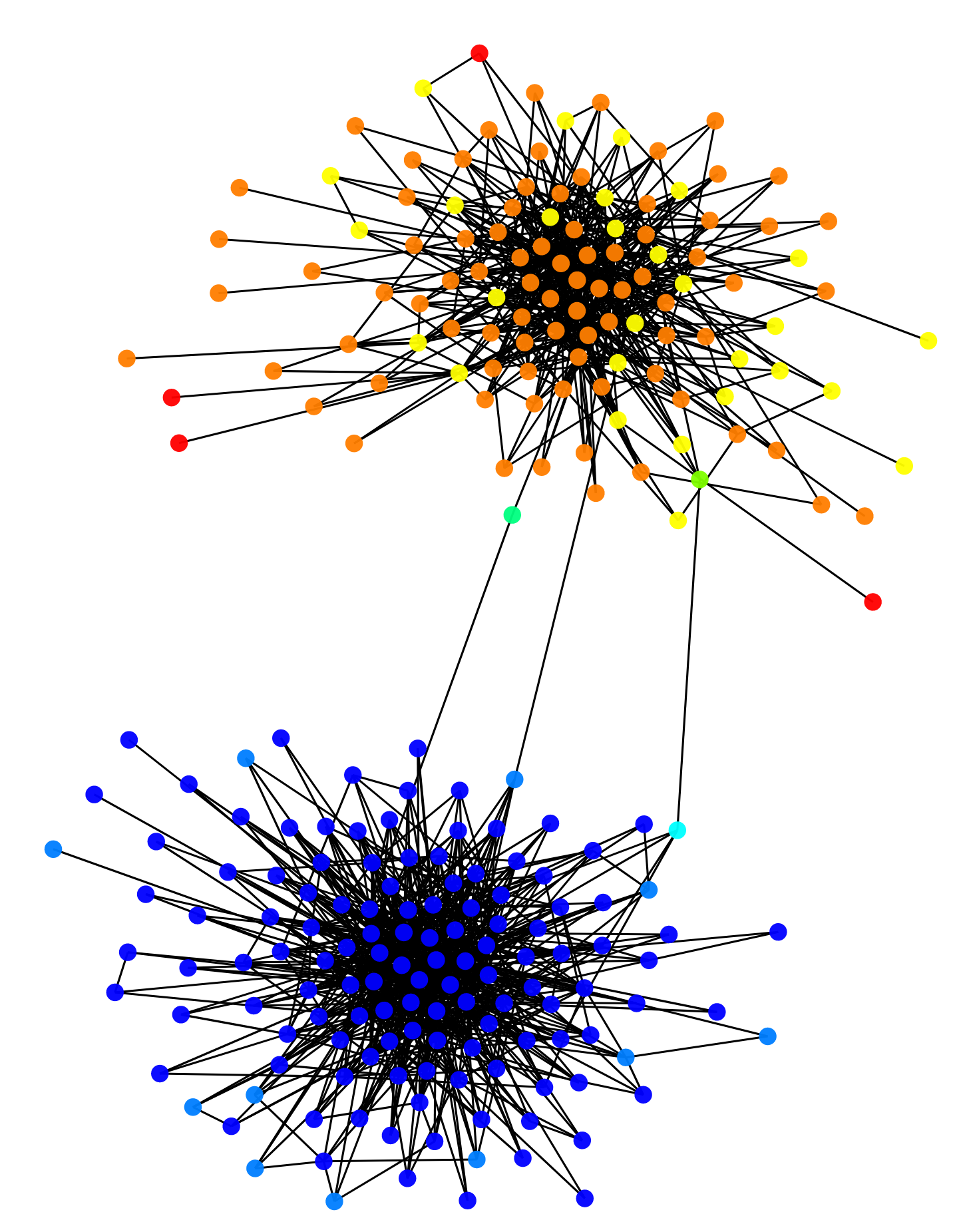} 
\label{fig:feature-diffusion-example-X5}
}

\caption{Feature diffusion example via $\mD^{-1}\mA$ in a barbell graph (top) and a graph following the Block Chung-Lu model (bottom). 
For demonstration/visualization purposes, we use a single feature, the value of which is denoted by the node color. 
Feature values were drawn from the uniform distribution on the open interval $(0,1)$. For the Block Chung-Lu model, the graph was generated with $\exp(1.7)$.
See text for discussion.
}
\label{fig:feature-diffusion-example}
\end{figure}

In Figure~\ref{fig:feature-diffusion-example}, we provide an illustration of Theorem~\ref{thm:feature-diffusion-conv} for further intuition.
This example clearly shows why feature diffusion gives rise to community-based embeddings.
In particular, after only 3 iterations of feature diffusion, the diffused features of nodes in either community become indistinguishable from one another.
More precisely, the diffused feature values of nodes within the same community are identical to one another after only a few iterations.
Furthermore, even after the first iteration of feature diffusion (Figure~\ref{fig:feature-diffusion-example-X1}), the features of nodes in the same community appear more similar and after only 3 iterations, the features of nodes in each community are indistinguishable from one another (Figure~\ref{fig:feature-diffusion-example-X3}).

Graph diffusion has also been used in community detection methods for decades~\cite{barbieri2013cascade, kloster2014heat, lin2015understanding}.
The diffusion process adopted by GCN is based on the Laplacian which has traditionally been used for community detection~\cite{graph-clustering-survey}.
Furthermore, recent work has also shown that embeddings from GCN are useful for community detection~\cite{bruna2017community,shchur2018overlapping,chen2018supervised}. 
The Laplacian-based diffusion used by GCN essentially uses a weighted sum to aggregate features from a node's neighbors. 

While the diffusion process is dependent on walks between node pairs, methods falling into this category do not explicitly leverage walks to approximate the graph structure. 
Instead, they propagate features over the graph structure up to $t$-orders, and characterize individual nodes by collecting the diffused feature values in the neighborhood~\cite{xiang2018feature}.
Thus, embeddings based on feature diffusions are community-based as the diffusion process is fundamentally tied to proximity in the graph as opposed to structural properties of nodes.

While all work described above essentially uses sum as a diffusion operator, 
DeepGL proposed the idea of using general aggregation functions.
Intuitively, DeepGL~\cite{deepGL} replaces the sum aggregator (which is naturally represented by a matrix-vector or matrix-matrix multiplication) with a general aggregation function $\phi$.
More generally, unlike previous work that used only sum, DeepGL uses multiple aggregation functions.
Examples of $\phi$ include min, max, product, mean, median, mode, $L_1$, $L_2$, RBF or more generally, any function that can be defined between a node $i$ and its neighborhood (or $k$-hop neighborhood).
More recently, this idea has been adopted in other works such as GCN-GraphSage~\cite{graphsage} and MultiLENS~\cite{multilens}.
However, replacing sum with a different aggregation function (or even multiple aggregation functions) does not change the fact that these methods are community-based in general for large $k$.

There are a few cases where feature diffusion can be used to derive role-based embeddings.
Note that when $\mX$ is motif/graphlet features and the number of feature propagations $k$ is 1,
then role-based embeddings can be derived.
Intuitively, the motif/graphlet features are not smoothed out for $k=1$.
However, as $k$ increases, the impact of the motif/graphlet features in their ability to capture the structural features are lost since they become increasingly similar to their neighbors.
Clearly, role-based embeddings can also be derived if $\mX$ representing motif/graphlet features is used without any diffusion (degenerate case).

\section{Role-based Embedding} \label{sec:role-based-embeddings}
This section discusses the main mechanisms behind role-based embeddings.
One notable observation is that the general mechanisms behind role-based embeddings all exploit an initial (small) set of structural features in some fashion.
As such, these mechanisms give rise to feature-based role methods~\cite{roles2015-tkde} where nodes with similar network neighborhoods will have similar embeddings when such mechanisms are used, despite that the nodes may be in different parts of the graph or even different graphs altogether.
We summarize the role-based embedding mechanisms in Table~\ref{table:embedding-methods} along with a few representative methods that use each mechanism.

\subsection{Graphlets} \label{sec:role-based-embeddings-graphlets}
We first give the definition of graphlets (network motifs/induced subgraphs) and orbits, then show how they can be used for learning role-based embeddings.
\begin{Definition}[{\sc Graphlet}]\label{def:graphlet}
A $k$-vertex graphlet $H = (V_k,E_k)$ is an induced subgraph consisting of a subset $V_k \subset V$ of $k$ vertices from $G = (V,E)$ together with all edges whose endpoints are both in this subset $E_k = \{ \forall e \in E \,|\, e = (u,v) \wedge u,v \in V_k \}$.
\end{Definition}\noindent
The edges of a graphlet can be partitioned into a set of automorphism groups called orbits based on the \emph{position} (or ``role'') of an edge in a graphlet~\cite{prvzulj2007biological,pgd}.
Formally,
\begin{Definition}[{\sc Orbit}]\label{def:orbit}
An automorphism of a $k$-node graphlet $H_t = (V_k,E_k)$ is defined as a permutation of the nodes in $H_t$ that preserves edges and non-edges.
The automorphisms of $H_t$ form an automorphism group denoted as $Aut(H_t)$.
A set of nodes $V_k$ of graphlet $H_t$ define an \emph{orbit} iff 
(i) for any node $u \in V_k$ and any automorphism $\pi$ of $H_t$, $u \in V_k \Longleftrightarrow \pi(u) \in V_k$; and
(ii) if $v,u \in V_k$ then there exists an automorphism $\pi$ of $H_t$ and a $\gamma > 0$ such that $\pi^{\gamma}(u)=v$.
\end{Definition}\noindent

\begin{figure}[h!]
\centering
\includegraphics[width=0.40\linewidth]{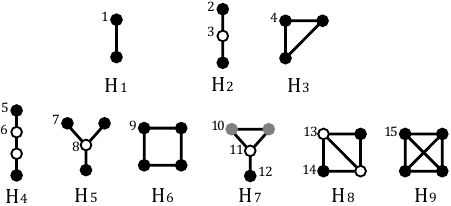}
\vspace{-2mm}
\caption{All $9$ graphlets and $15$ node orbits with $\{2,3,4\}$-nodes.
Each unique node position of a graphlet is labeled, \eg, nodes in the 4-star graphlet ($H_5$) have two unique positions, namely, the star-center (hub) position or the star-edge (peripheral) position.
}
\label{fig:all-conn-node-orbits-up-to-4-nodes}
\end{figure}

Graphlets naturally capture the key structural properties of edges and nodes in the graph as shown in Figure~\ref{fig:all-conn-node-orbits-up-to-4-nodes}.
In particular, Figure~\ref{fig:all-conn-node-orbits-up-to-4-nodes} shows the full spectrum of connected graphlets with $\{2,3,4\}$-nodes; each set of 
$k$-node graphlets are ordered from least to most dense.
Notice that graphlets are the fundamental building blocks of graphs since any graph (or $\ell$-hop neighborhood subgraph surrounding a node/edge) can be decomposed into its smaller subgraph patterns (graphlets).
In other words, any (sub)graph can be represented using only $k$-node graphlets.
Therefore, by definition, the graphlets and their counts must capture the structural properties that are important to a node or edge.
As such, graphlets (and their edge orbits) capture precisely the notion of role.
This can be trivially verified from Figure~\ref{fig:all-conn-node-orbits-up-to-4-nodes} as many of the individual graphlets can even capture the traditional examples of roles used in the literature. 
Recall roles represent node (or edge~\cite{ahmed2017roles})
connectivity patterns such as hub/star-center nodes, star-edge nodes, near-cliques or bridge nodes connecting different regions of the graph.
Graphlets capture the full spectrum of possible connectivity/subgraph patterns arising in graphs as shown in Figure~\ref{fig:all-conn-node-orbits-up-to-4-nodes}, which lies at the heart of the notion of roles (Section~\ref{sec:roles}).
Intuitively, two nodes belong to the same role if they are structurally similar with respect to their general connectivity/subgraph patterns~\cite{roles2015-tkde}.
Therefore, it is only natural to consider graphlet features when learning role-based embeddings.
There is a broad class of embedding methods that represents structural information using graphlets to learn role-based embeddings.
Graphlets and the statistics (\eg, frequencies) 
carry significant information about the structural properties of nodes and edges and have been used for many applications~\cite{holland1976local,faust2010puzzle,milo2002network,ahmed2017graphlet}.
The set of decomposed graphlets $\{H_1, H_2, \cdots H_d\}$ can be used to characterize both the whole graph and individual nodes/edges by counting the number of times in the embedded $d$-dimensional vector. 
Intuitively, nodes associated with similar graphlet/motif types (\eg, triangle, star) and counts are structurally similar and thus are embedded closer in some low-dimensional space.

The graphlet features are computed for each node (or edge) in the graph and naturally generalize across graphs (for transfer learning tasks) since they represent ``structural graph functions'' that are easily computed on any arbitrary graph.
Fast algorithms for counting such graphlets/network motifs in very large graphs have become common place,
\eg, PGD~\cite{pgd} takes a few seconds to count graphlets in very large networks with hundreds of millions of edges.
Furthermore, since most embedding methods are not exact 
and we typically only care about the relative/approximate magnitude of the count (\eg, whether the count is on the order of $10^1$, $10^2$, $10^3$ and so on), and not the actual exact count, we can also leverage provably accurate graphlet estimators to obtain graphlet counts even faster.

\begin{figure}[t!]
\centering
\hspace{-5mm}
\subfigure[Initial graph $\mA$]{
\includegraphics[width=0.28\linewidth]{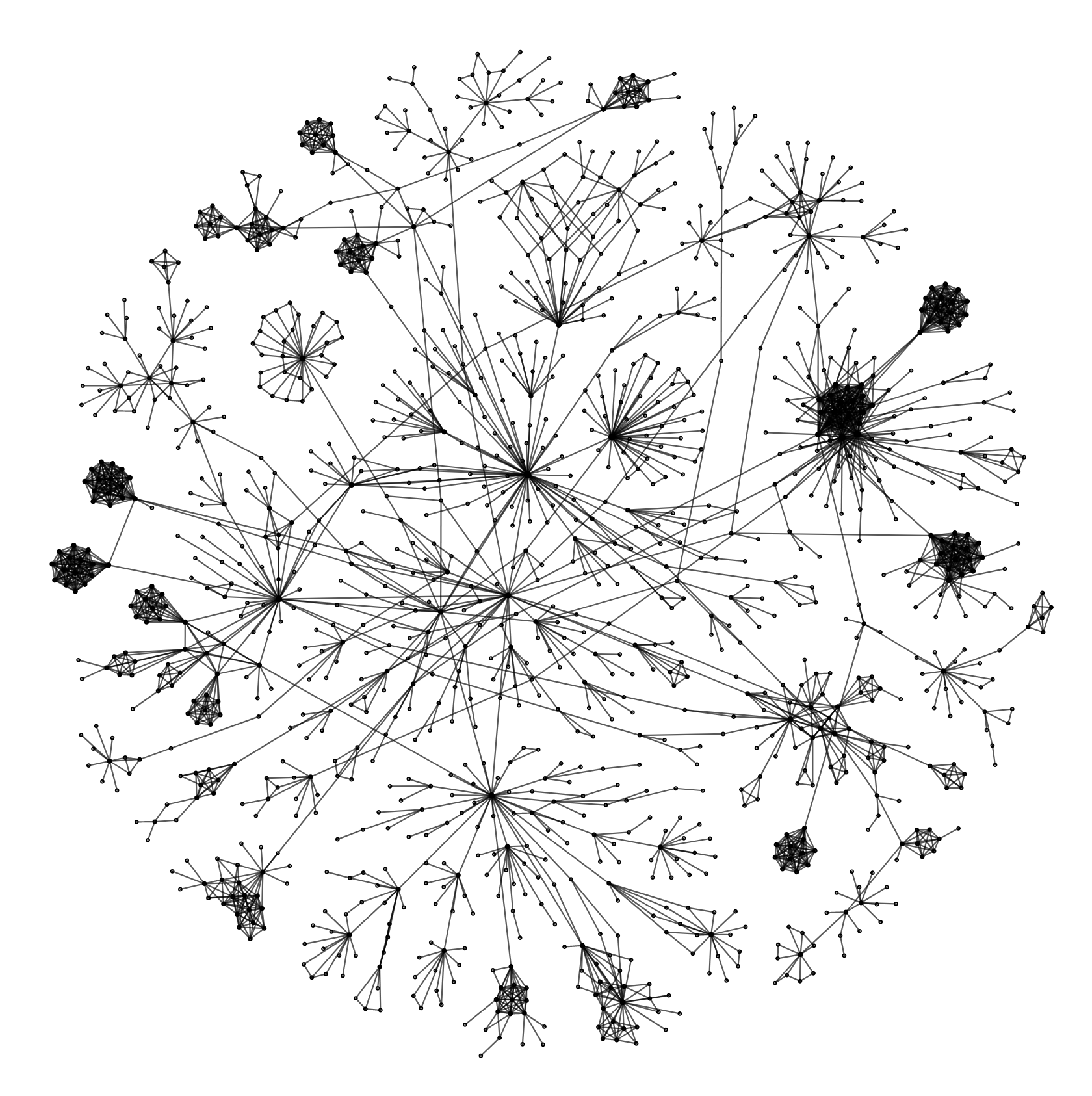} 
\label{fig:web-google-initial-graph}
}
\hspace{-3mm}
\subfigure[Weighted 4-clique graph
{\protect\includegraphics[width=2.4mm]{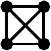}}
]{\includegraphics[width=0.24\linewidth]{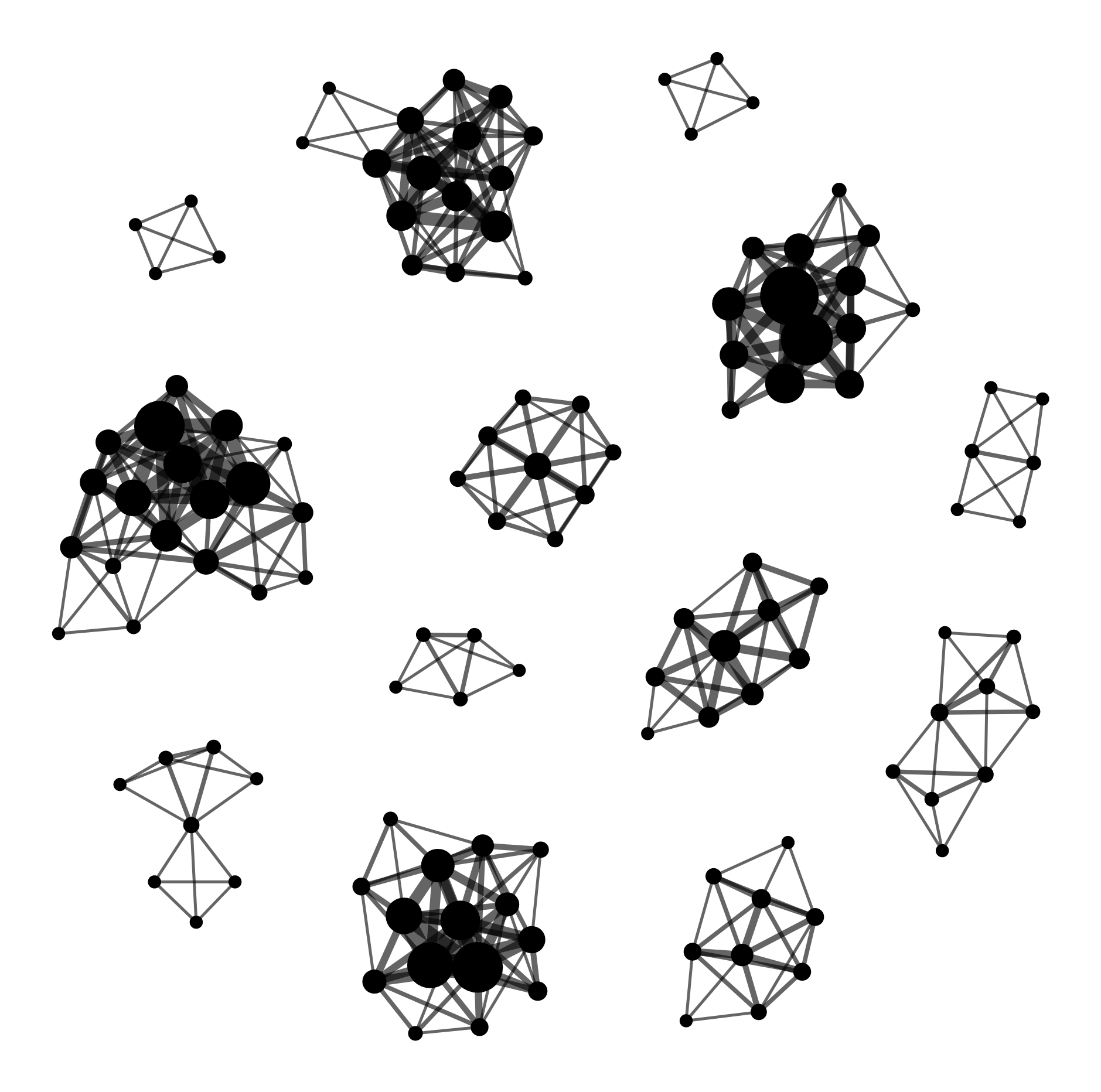}
\label{fig:motif-graph-comparison-web-google-4clique}
}
\hfill
\subfigure[Wt. 4-cycle graph {\protect\includegraphics[width=2.4mm]{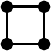}}]
{\includegraphics[width=0.24\linewidth]{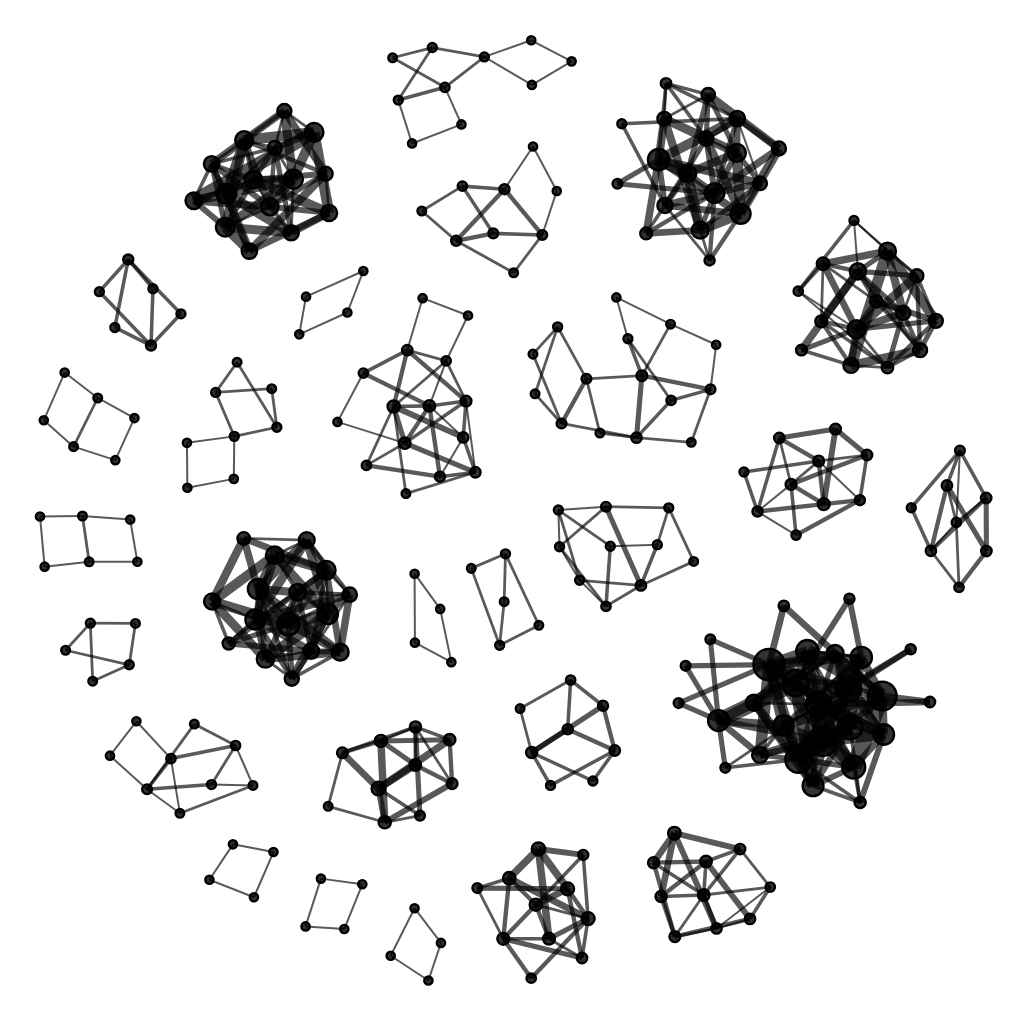}
\label{fig:motif-graph-comparison-web-google-4cycle}
}
\hfill
\subfigure[Wt. 4-star graph {\protect\includegraphics[width=2.4mm]{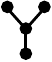}}
]{\includegraphics[width=0.24\linewidth]{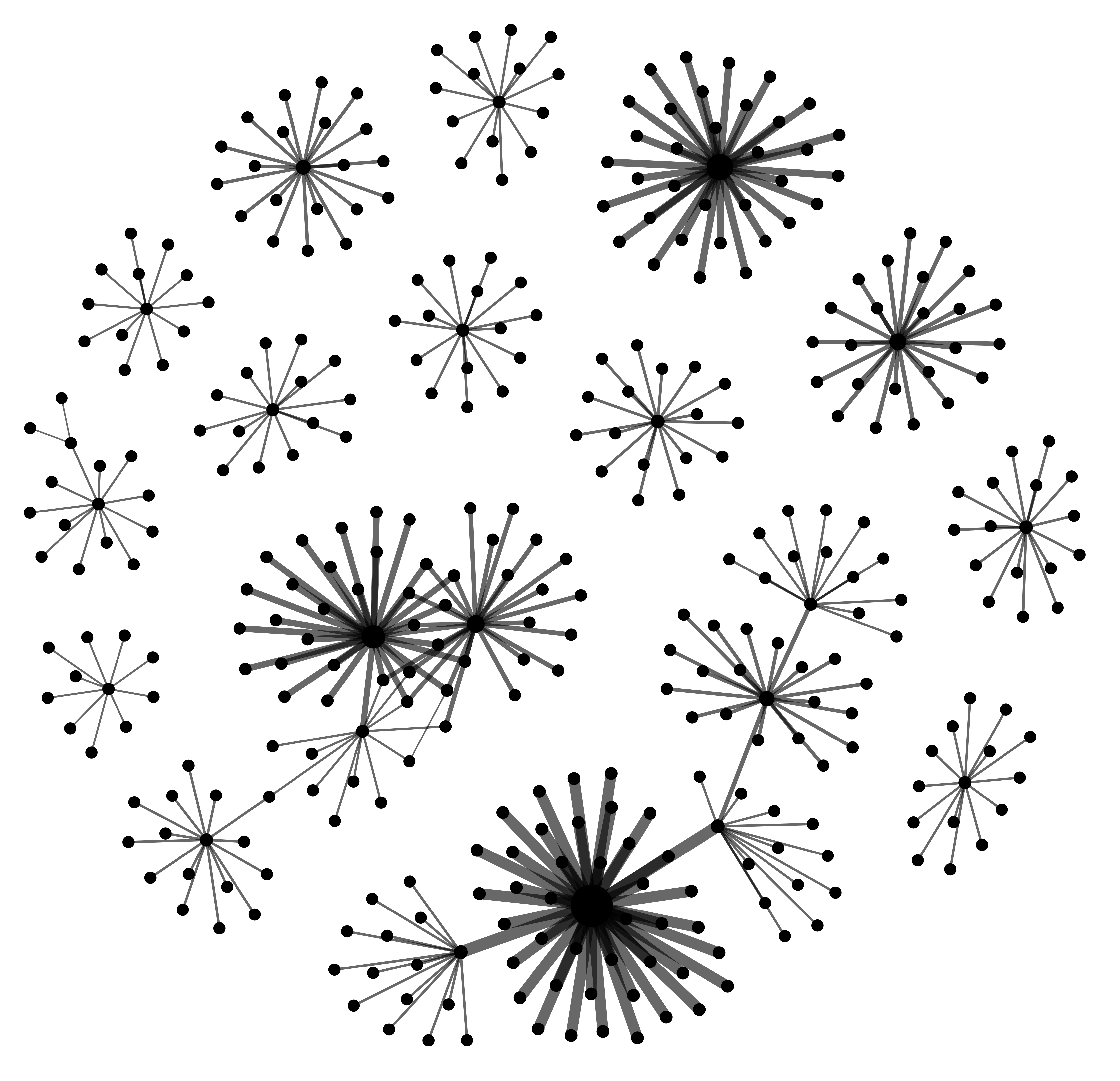}
\label{fig:motif-graph-comparison-web-google-4star}
}
\hspace{-2mm}

\caption{
Graphlet/motif graphs differ in structure \textit{and} weight.
Size (weight) of nodes and edges in the motif graphs correspond to the frequency of the motif.
In this example (web-google), the initial (edge motif) graph 
is fractured into many 
disconnected components when deriving the motif graphs.
This is due to the constraint that each edge in an arbitrary motif graph contain at least a single motif. 
Edges are removed if they do not participate in at least one
4-clique (b), 
4-cycle (c), 
or 4-star (d). 
}
\label{fig:motif-graph-comparison-web-google}
\end{figure}

Representative approaches falling into this category are as follows. 
Role2vec~\cite{role2vec} proposed an end-to-end inductive framework to learn role embeddings that capture the structural similarity among nodes and generalizes across networks. 
To achieve this, role2vec introduced the general notion of \emph{feature-based random walks}, to replace the traditional random walks (\ie, random sequence of node ids) by walks that represent the structural similarity among nodes, where each walk is a sequence of features or functions of multiple features.
Thus, the feature-based walks find nodes with similar structure identified by structural properties and higher-order features (\eg, graphlets), and enable  space-efficient role embeddings. DeepGL~\cite{deepGL} learns inductive graph functions where each represent a composition of relational operators/aggregator functions applied to a graphlet/motif feature.
HONE~\cite{HONE} proposed the notion of higher-order network embeddings and described a framework based on weighted k-step motif graphs to learn the low-dimensional role-based embeddings.
More recently, higher-order motif-based GCN's called MCN were proposed by~\citet{lee18-higher-order-GCNs}.
MCN leverages the weighted motif-based matrix functions introduced in HONE~\cite{HONE} to learn role-based embeddings.
While HONE and MCN (a higher-order generalization of GCN) also use different forms of feature diffusion, they are nevertheless role-based since they do not leverage $\mA$ directly, but instead use $\mA$ to derive a set of \emph{weighted motif graphs} (\ie, weighted motif adjacency matrices $\mW_1, \mW_2, \ldots, \mW_k$) that are then used to derive node embeddings.
In particular, given a motif $H$, the weighted motif adjacency matrix of $H$ denoted $\mW_H$ is defined as:
\begin{equation} \label{eq:weighted-motif-graph}
(\mW_H)_{ij} = \text{number of instances of motif } H \text{ that contain nodes } i \text{ and } j
\end{equation}
The weighted motif adjacency matrices differ fundamentally in structure (and weight) when compared to the original graph as shown in Figure~\ref{fig:motif-graph-comparison-web-google}.
In particular, the motif graphs typically consist of many connected components, \ie, the graph shatters into many connected components due to the requirement that each edge have at least $\delta>0$ motifs.
This requirement acts as a filter, removing many of the unimportant edges, and highlighting only the edges (and structures) in the graph that contain at least one such occurrence of $H$ (or more generally $\delta$ occurrences of $H$).
In Figure~\ref{fig:motif-graph-comparison-web-google}, the motif graphs immediately reveal nodes with similar structural properties.
For instance, the weighted 4-star graph (Figure~\ref{fig:motif-graph-comparison-web-google-4star}) fractures the graph into many connected components where each connected component consists of nodes that are either 
(i) \emph{star-center} (hub) nodes of some larger star structure or 
(ii) \emph{star-edge} (peripheral) nodes.\footnote{Recall from Section~\ref{sec:comm-and-roles-prelim} that star-centers (hubs) and star-edges are two of the classic examples used for roles.}
Furthermore, the graphlet/motif graphs immediately reveal larger subgraph patterns, \eg, the 4-star graph shown in Figure~\ref{fig:motif-graph-comparison-web-google-4star} shows large stars made up of many 4-stars.
Both of these properties are important when considering feature diffusion (via a graph smoothing operator such as the normalized Laplacian $(\eye - \mD^{-\frac{1}{2}}\mA\mD^{-\frac{1}{2}})\mX$ or $(\mD^{-1}\mA)\mX$) and its meaning and impact when used with these weighted motif graphs.
As such, the diffusion process is performed over each connected component in a motif graph, and has less of an impact since all nodes in the motif graph by definition have similar structural properties.

\subsection{\emph{Feature-based} Walks} \label{sec:feature-based-walks}
While we theoretically showed in Section~\ref{sec:node-edge-comm-embedding} that walk-based embedding methods derive community / proximity-based embeddings, \citet{ahmed17Gen-Deep-Graph-Learning} proposed the notion of a feature-based walk (attributed random walk) that allows existing walk-based methods to be generalized for learning role-based embeddings.
The general idea is to generate walks that represent sequences of feature values as opposed to sequences of node ids as done in DeepWalk, node2vec, among many others.
Intuitively, the feature values in the walks naturally generalize across graphs since they can represent general graph functions (like degree, number of triangles, 4-node cycles).
By applying a mapping function $\Phi$ to map sequences of node ids (walks) to their associated feature-values, the skip-gram model (or any other model that uses the walks) would preserve the similarity in terms of attributes in the embeddings. 
Formally, the walk consisting of node feature-values/types/labels/attributes is defined as follows:
\begin{Definition}[Feature-based Walk]\label{def:feature-based-walk}
Let $\mathbf{x}_i$ be a $K$-dimensional feature vector for vertex $v_i$. 
A feature-based (or attributed) walk of length $L$ is a sequence of adjacent feature-values, 
\begin{equation}
\Phi(\vx_1), \Phi(\vx_2), \cdots, \Phi(\vx_L)
\end{equation}
induced by a sequence of indices $(v_1,v_2, \cdots, v_L)$ generated by a random walk of length L starting at $v_1$, and a function $\Phi$ that maps a feature vector $\vx$ to a type $\Phi(\vx)$.
\end{Definition}
Definition~\ref{def:feature-based-walk} can be used in a variety of ways and gives rise to many interesting role-based methods~\cite{role2vec}.
For instance, instead of using a mapping function (which can be thought of as replacing $\Phi$ with the identity function), one can simply use one or more features to derive feature-based walks for each different feature.

Recently, an approach called role2vec~\cite{role2vec} was proposed that learns role-based node embeddings by first mapping each node to a type (or role) via a function and then uses the notion of attributed random walks to derive role-based embeddings for the nodes that capture structural similarity.
Since the ``feature-based'' random walks by definition capture the structural properties (the features can be thought of as describing the topological/structural characteristics of a node), node embeddings (representations, encodings) learned from these attributed/feature-based walks are able to capture roles as opposed to communities.
Attributed random walks~\cite{ahmed17Gen-Deep-Graph-Learning,role2vec} have since become increasingly popular with applications in entity resolution~\cite{node2bits}, churn prediction~\cite{liu2019micro,liu2018semi}, among many others~\cite{al2018learning,rozemberczki2019multi,zhang2019attribute,attr-rand-walk-matrix-neurips20}.
In particular, node2bits~\cite{node2bits} builds on the idea of feature-based walks and also the notion of temporal walks that obey time from CTDNE~\cite{nguyen2018continuous} in order to obtain temporal, feature-based context around the nodes (based on both structural features and node attributes). It then aggregates the contexts into node-level histograms and obtains compact, binary node representations that preserve context similarity via locality sensitive hashing (LSH) and uses the final embeddings for the application of visitor stitching/entity resolution.
More recently,~\cite{liu2019micro,liu2018semi} proposed SimSum that leverages attributed random walks to analyze churn in mobile games at both the micro and macro levels.
Another extension of role2vec called RiWalk~\cite{riwalk} introduced a new role mapping function based on the shortest path and Weisfeiler-Lehman kernels.
Many other works have also used the notion of attributed random walks for a variety of applications and problem settings~\cite{al2018learning,rozemberczki2019multi,huang2019graph,zhang2019attribute}.

\subsection{\emph{Feature-based} Matrix Factorization} \label{sec:feature-based-matrix-factorization}
There are also role-based structural embeddings that use a form of matrix factorization over a matrix of structural features~\cite{roles2015-tkde}.
This is in contrast to community-based embeddings that use matrix factorization over the graph's adjacency matrix.
This class of role-based embeddings largely depends on the structural feature matrix used in the factorization.
For instance, suppose the features in the matrix are all correlated with communities (as opposed to roles; and thus they are not \emph{structural features}), then the resulting embeddings would in fact be community-based and not role-based.
Thus, the most critical step in these methods is to ensure the initial set of structural features are appropriate and are most suitable for capturing roles.

This class of feature-based role embeddings along with a general framework for computing them was introduced in~\cite{roles2015-tkde}.
One such work by~\citet{rolX} starts with degree/egonet-based features, aggregates them recursively, and then uses Non-negative Matrix Factorization (NMF) over the feature matrix to derive roles.
More recently, \citet{rossi2013dbmm-wsdm} proposed modeling feature-based roles in dynamic networks whereas~\citet{gilpin2013guided} used a sparsity regularized NMF (as well as other convex constraints) to learn better roles.
While all previous methods directly learn node embeddings, \citet{ahmed2017roles} learns role-based \emph{edge} embeddings.
This approach starts with higher-order graphlet features that explicitly captures the notion of roles (see Section~\ref{sec:role-based-embeddings-graphlets}) and iteratively computes additional higher-order features via relational aggregates over the neighborhood, and then factorizes this matrix of structural features to obtain role-based embeddings of the edges.
More recently, xNetMF~\cite{heimann2018regal} computes degree-based features from a node's neighborhood at different hops, and uses implicit matrix factorization over this feature matrix to obtain generalizable embeddings that are suitable for network alignment.
EMBER~\cite{ember} generalizes xNetMF to weighted and directed graphs by extending the degree-based distributions to weighted distributions defined over directed neighborhood contexts. 
SEGK~\cite{segk} also computes the similarity between nodes by leveraging different hops of neighborhoods, but utilizes graph kernels for the comparison, and then factorizes the resultant kernel matrix to obtain the structural node embeddings.
While most of the approaches discussed in this section so far first compute a tall-and-skinny ``structural'' feature matrix, struc2vec~\cite{struc2vec} computes multiple (large and dense) node-by-node feature matrices between all pairs of nodes using dynamic time warping (DTW) distance based on sequences of node degrees (\ie, degrees of the neighbors of a node).  
Afterwards, explicit walks over the resultant dense multi-layer graph are sampled and used to derive embeddings in a similar fashion to DeepWalk.

Moreover, we note that any of the previous structural embeddings from Sections~\ref{sec:role-based-embeddings-graphlets}-\ref{sec:feature-based-walks} can be used as input into matrix factorization to learn more compact and space-efficient role-based embeddings~\cite{roles2015-tkde}.

\section{Applications} \label{sec:applications}
In this section, we describe applications for community (proximity) and role-based embeddings.
For each application, we discuss conditions including data characteristics, noise, and variants of the applications where community and role-based embeddings are most suitable.
Notably, community or role-based embeddings are shown to be useful for the same applications such as classification, link prediction, and anomaly detection. 
The fundamental difference of whether community or role-based embeddings are preferred depends entirely on the underlying data characteristics (\eg, homophily vs. heterophily, noisy/missing data vs. clean/accurate data) and problem setting/assumptions.

\subsection{Node classification} \label{sec:app-classification}
Embeddings have been used to improve node classification performance.
We discuss a few different node classification tasks below and mention the key differences and data characteristics that make community-based or role-based embeddings more appropriate.

\subsubsection{Community-based embeddings}
Semi-supervised classification in graphs 
typically performs best with community-based embeddings since these methods iteratively predict labels of neighbors and propagate them to neighboring nodes~\cite{sen2008collective}.
In other words, the labels of neighboring nodes are repeatedly diffused to the neighbors until convergence.
The overall process is similar to feature diffusion-based methods from Section~\ref{sec:node-edge-embedding-comm-feature-diffusion}.
Typically, these methods assume a small fraction of nodes with known labels are given (for training), and since neighboring nodes are assumed to be labeled the same, these methods are most useful for graphs with significant homophily/large relational autocorrelation, \ie, graphs where the node labels are highly correlated with their immediate neighbors~\cite{homophily-LaFond2010,neville:srl04,Koutra17-book}.
Examples of such graphs with strong homophily include Cora, CiteSeer, among many others~\cite{mcdowell2009cautious}.
As such, if such \emph{strong homophily} exists, then community-based embedding methods are most appropriate.
This is the reason why many community-based embedding methods such as GCN~\cite{gcn} are evaluated for semi-supervised classification using graphs with \emph{strong homophily} such as Cora, CiteSeer, PubMed, and others.
Nevertheless, community-based embeddings are also preferred for general node classification with homophily~\cite{deepwalk,node2vec,line,ComE}, going beyond the semi-supervised classification setting.

\subsubsection{Role-based embeddings}
Role-based embeddings are based on \textbf{structural} similarity (Definition~\ref{def:feature-based-struct-sim}) and thus appropriate for classifying nodes with similar functionality (roles) in terms of their structural properties, \eg, triangles, betweenness, stars, etc. 
For collective/semi-supervised classification, there are some instances where role-based embeddings can perform better than community-based. 
For instance, role-based embeddings are often useful for graphs with weak/low homophily.
Such graphs may have weak homophily due to noise, incompleteness, or other data collection/sampling issues,
or graphs with heterophily where node labels (and attributes) are not correlated with the labels of their neighbors~\cite{rogers1970homophily,GatterbauerGKF15,peel2017graph,rossi18-rsm-bigdata}.
For instance, molecular, chemical, and protein networks often have between 2 and 20 class labels, which are highly correlated with the structural properties (\eg, graphlets/network motifs) and behavior surrounding a given node or edge in the graph~\cite{vishy2010graphKernels,GardinerWA:2000}. 
Networks of email communication in the workplace is another example where the professional roles of nodes (e.g., C-suite employees vs.\ managers) correlate with their structural properties in the network~\cite{ember}. 
In these cases, the nodes whom share class labels are often not directly connected, or even in the same community, but share similar structural properties and behavior (or role/position) in the network~\cite{roles2015-tkde}.

While community-based embeddings are primarily useful for semi-supervised classification, role-based embeddings are also well-suited for \emph{across-network (relational) classification} where the goal is to learn a classification model on one graph and then use it to predict the labels of nodes in an entirely different graph that may not share any of the same nodes.
The two graphs could have completely different nodes (\ie, node ids) or may have some nodes in common between the two graphs, \eg, in temporal networks where there is a sequence of graphs over time.
This application is sometimes called relational classification as opposed to 
semi-supervised classification, see~\cite{rossi12jair} for more details.

\subsection{Link prediction} \label{sec:app-link-pred}
Link prediction is another important application where embeddings can be used to improve performance over simpler approaches such as common neighbors, Jaccard similarity and the ilk~\cite{node2vec,kipf2016variational,role2vec}.
Given a graph $G=(V,E)$, the link prediction task is to predict a set of (top-$k$) missing (unobserved) or future links $E^{\prime}$ such that $E^{\prime} \cap E = \emptyset$.
Given node embeddings (either community-based or role-based), links can be predicted by computing edge feature vectors $g(\vx_i, \vx_j), \forall i,j$ pairs (\ie, in the training set), and then learning a model based on these, which is then used to predict the likelihood that a link exists between any arbitrary pair of nodes.
Alternatively, we can directly compute a score without learning a model.

\subsubsection{Community-based Embeddings}
In many cases, the missing or future links are assumed to arise between nodes that share many of the same neighbors. 
Hence, given nodes $i$ and $j$ such that $(i,j) \not\in E$, community-based embeddings are useful when $|N_{i} \cap N_{j}|>0$, that is, $i$ and $j$ share at least one neighbor (1-hop away).
The above condition implies that $i$ and $j$ are near one another in the graph due to the sharing of at least one neighbor among them.
We call such predicted links short-range, since they are between nodes that are close to one another in the graph.
It is because of this property that community-based embeddings will work best for predicting such links (especially if the pair of nodes are both in the same community, and therefore will be embedded in a similar fashion)~\cite{node2vec,gcn}.

\subsubsection{Role-based Embeddings}
There are also many settings and applications where role-based embeddings perform best for link prediction.
In some cases, the graph data may be noisy or incomplete due to sampling or data collection issues~\cite{ahmed2014network}, and therefore the actual links may not be close in terms of graph distance.
More formally, $|N_{i} \cap N_{j}|=0$.
In fact, the actual links could be between nodes that are far from one another in the graph (long-range) or even in different connected components~\cite{gilpin2013guided,node2bits,role2vec}. 
We call such links long-range as opposed to short-range.
Furthermore, links may also be predicted between nodes far away in the graph to improve relational autocorrelation or similarity, see~\cite{rossi12jair,ghost-edges,LGM,lassez2008ranking}.

\subsection{Graph alignment and classification}
Node embeddings have also been used for graph-level tasks, such as graph alignment and classification. 
Network alignment seeks to find the corresponding nodes across two or more networks. 
It can be thought of as a link prediction problem, where the links are predicted between two nodes $i$ and $j$ such that $i$ is in one graph $G$ and $j$ is a node in another graph $G^{\prime}$.  
Graph classification aims to categorize graphs into classes based on their structure. 
We discuss the properties that make community-based or role-based embeddings appropriate for these tasks.

\subsubsection{Community-based Embeddings} 
Methods that learn proximity-based node embeddings via diffusion or propagation, such as GCN~\cite{gcn} and GraphSAGE~\cite{graphsage}, can be used to obtain \textit{supervised} network representations by aggregating (e.g., concatenation) the node embeddings. Supervision makes these representations suitable for graph-level tasks such as network classification. Some graph neural network-based methods tailored to graph classification supervise layer-wise node pooling to learn expressive, hierarchical network representations~\cite{ying2018hierarchical}, or supervise node grouping (or graph summarization) in order to learn more robust-to-noise network representations, provide interpretability, and achieve better scalability~\cite{yan2019groupinn}.
Transductive community-based embeddings (that do not leverage structural or other node features) are specific to a network, and fail to tackle cross-network tasks such as unsupervised network alignment~\cite{HeimannK17-mlg} or unsupervised network classification (e.g., by simply aggregating unsupervised community-based node embeddings into a network-level vector representation). 
However, recent work inspired by machine translation has shown that transductive proximity-based embeddings can be used to improve the performance in network alignment by encouraging the alignment of local neighborhoods across different graphs (rather than greedily matching nodes with the same structural properties), but \textit{only after} appropriately aligning/rotating the embedding spaces of the networks~\cite{ChenHVK20}.

\subsubsection{Role-based Embeddings}
Role-based embeddings generalize across networks~\cite{roles2015-tkde}, so they are useful in network alignment and identity resolution (or user stitching, which can be seen as a node correspondence problem within a single network or across multiple networks)~\cite{zhang2013predicting,heimann2018regal,node2bits,gilpin2013guided}.
In such applications, transductive community-based embeddings are unable to be used off-the-shelf since 
the corresponding embeddings belong to different, \textit{unaligned} latent spaces~\cite{HeimannK17-mlg}.
However, since role-based node embeddings are based on structural properties (\eg, degree, graphlet counts, betweenness) that generalize over any graph, they can naturally be used in such settings, referred to as graph-based transfer learning~\cite{deepGL}. Role-based embeddings also generalize across different parts of a single network, making them suitable for matching nodes corresponding to the same entity (i.e., identity resolution or user stitching)~\cite{node2bits}. 
Beyond alignment and identity resolution, aggregating role-based node embeddings can provide a powerful descriptor of an entire network, which is useful for network classification. 
For example, a graph descriptor can be obtained by using role-based embeddings (\eg, inductive extension of xNetMF~\cite{heimann2018regal,HeimannSK19}, SEGK~\cite{segk}) and creating a graph-level feature vector based on the distribution or spatial overlap of the embeddings~\cite{HeimannSK19,embed-graphsim} or a graph kernel~\cite{segk}. 
On the other hand, embeddings based on proximity are not suitable for this setting.

\subsection{Anomaly detection}
Community or role-based embeddings also have applications in graph-based anomaly detection where the goal is to identify (node/edge/subgraph) anomalies that do not conform to the expected behavior in the graph~\cite{akoglu2015graph,Fond2018DesigningSC,abello2010detecting}.
Such nodes/edges/subgraphs with non-conforming behavior are known as anomalies, outliers, or exceptions~\cite{chandola2009anomaly}.
This problem also has connections to change detection in temporal networks.
There are specific problem settings in anomaly detection where community-based or role-based embeddings are more appropriate.
We discuss these settings below.

\subsubsection{Community-based Embeddings}
One anomaly detection application of community-based embeddings is in the detection of anomalous global changes between different static snapshot graphs derived from the temporal network (sequence of edge timestamps)~\cite{ide2004eigenspace,multilens}.
One particular instance of this problem uses communities (groups of nodes that are tightly/densely connected) and considers an anomaly (or change-point) to occur when the nodes and their community-based embeddings differ significantly from the previous time at $t-1$~\cite{ide2004eigenspace,chen2012community}.
The underlying assumption is that the communities and community-based embeddings will remain largely stationary over time, \ie, there is minor differences between time $t-1$ and time $t$~\cite{akoglu2015graph}.
Thus, when a group of nodes suddenly becomes more similar to another community, a flag is raised and a change-point is detected~\cite{sun2007graphscope}.

\subsubsection{Role-based Embeddings}
There are many applications and problem settings where role-based embeddings are more useful for graph-based anomaly detection.
Role-based embeddings are often most useful for applications where anomalies can be defined
with respect to the structural properties and behavior in the network.
For instance, an anomaly in this setting might be when a node's structural properties/behavior 
differ significantly from all other nodes in the network~\cite{akoglu2015graph}.
Another slight variation of this problem can be defined for temporal networks as well.
In particular, suppose the goal is to detect nodes with sudden changes in their structural behavior.
In this example, node anomalies may represent users (or computers) that become infected with a virus/malware in the network and thus the nodes structural properties/behavior abruptly changes~\cite{rossi2013dbmm-wsdm,fu2009dynamic,dynamic-anomaly-survey}.

Many anomaly detection applications might benefit from the use of external knowledge (structural behavior/profile) that was found to be important in the detection of a new anomaly that has recently been detected in some other graph (\eg, an IP communication/traceroute network from another organization/company)~\cite{rolX}.
Role-based embeddings are therefore most useful for this application since they represent general structural properties important in the detection of the anomaly and the specific structural properties captured in the embedding can be transferred to another arbitrary network (and thus used as a signature) for detecting this new recent anomaly (\eg, the anomaly may represent a recent zero-day attack vector).

\subsection{Summarization / compression}
The overall goal of summarization/compression is to describe the input graph $G$ with a compact representation~\cite{liu2018graph,ahmed17streams}.
The precise way to do this fundamentally differs depending on whether communities or roles are preserved.

\subsubsection{Community-based Embeddings}
The majority of work in graph summarization are naturally based on community-based embeddings as they leverage the notion of communities directly.
Though there are various different summarization techniques, many methods leverage grouping or clustering, and represent each
group of densely connected nodes that are nearby one another (cluster or community) as a super-node and the edges between such nodes as super-edges~\cite{liu2018graph,koutra2014vog,shah2015timecrunch}. 
More recently, the notion of latent network summarization was introduced~\cite{multilens}; it leverages community-based embeddings by propagating structural features in the network via relational functions, and achieves compression by storing low-rank, size-independent structural feature matrices and the relational functions as the latent network summary.

\subsubsection{Role-based Embeddings}
Role discovery methods typically output a role graph (Definition~\ref{def:role-graph}) that succinctly represents the key structural roles and the dependencies between them~\cite{carrington2005models,roles2015-tkde}.
The role graph consists of super-nodes that represent roles and the edges between the roles are super-edges and encode the dependencies between the different roles.
The role graph represents a summary of the roles and relationships between the roles.
It can also be seen as a smaller model of the original graph and therefore can be viewed as a compressed representation of the overall graph as it succinctly represents the main structural patterns and the relationships between them (\eg, if a star-center node connects to many star-edge nodes, then the super-node that captures the star-center role will have an edge to the super-node representing the star-edge role).

\subsection{Visualization} \label{sec:visualization}
Community and role-based embeddings are also useful in visualization applications, especially as they relate to reducing information overload and in the visualization of large graphs~\cite{von2011visual,keim2008visual,abello2006ask}.
Recall that communities and roles are complimentary concepts (Table~\ref{table:roles-vs-comms}) and therefore both provide useful information when used to summarize the graph for visualization purposes.

\subsubsection{Community-based Embeddings}
In many visualization applications, community-based embeddings are useful when the graph/network data is too large to visualize all-at-once~\cite{newman2004finding,hu2015visualizing}.
In such problem settings, community-based embeddings can be used to derive communities which are then displayed to the user in the initial visualization of the graph.
This is used as a way to navigate large graphs and avoid the computational and visual problems that arise when visualizing large-scale graphs~\cite{von2011visual}.
The user can then visually select the community of interest, which is then displayed to the user.
In this example, once a community is selected, the user can view the nodes and edges that belong to it, while avoiding all other nodes and edges that are not of interest to the specific user/query~\cite{abello2006ask}.

\subsubsection{Role-based Embeddings}
In a similar fashion, role-based embeddings can be used for navigating large networks~\cite{graphvis,Koutra17-book}.
Suppose the user is only interested in hub (star-center) nodes, then we can immediately visualize all such nodes and their roles while avoiding the computational and visual issues that arise when trying to visualize large graphs. Work on vocabulary- or motif-based summarization of static or time-evolving graphs can also be leveraged in visualization of structural roles~\cite{koutra2014vog,DunneS13,ShahKJZGF17}. 
Other similar types of (structural role-based) queries and filtering can be performed to answer other questions of interest to the user.

\subsection{Clustering}
\label{sec:applications-clustering}
Community and role-based embeddings are also clearly useful in graph clustering (Section~\ref{sec:comm-and-roles-related-to-clustering}).
For simplicity, we have described roles and communities in Section~\ref{sec:comm-and-roles-related-to-clustering} with respect to hard assignments.
However, over a decade ago, methods for roles and communities that naturally output node embeddings have been investigated.
The node embeddings from these methods are sometimes referred to as node mixed-membership vectors.
In other words, community and role discovery are not different problems than node embeddings, since they both output node embeddings.

In this section, we discuss a few applications of communities and roles that make use of the hard assignments.
We also note that some of the previous applications leveraged the hard assignments.
However, we focus mainly on discussing other applications that have not yet been discussed above.

\subsubsection{Community-based Embeddings}
Given the community-based embeddings, we can use a clustering algorithm to derive hard cluster assignments. 
These hard cluster assignments have been useful for sampling-based applications.
In particular,~\citet{bilgic2010active} used them for relational active learning where nodes are actively sampled from every community. Hard cluster assignments can also be used to improve scalability of a variety of applications.
For instance, given a node embedding of interest, we can use the k cluster centroids to find the significantly smaller set (community cluster) of relevant nodes, without having to compare to all such nodes. 
Similar ideas may also be useful for improving search and recommendation systems~\cite{Sarwar02,li2003clustering}. In this context, similarity in representations can also be used to identify the top-$K$ related objects, without explicitly defining clusters. For instance, by learning propagation-based representations of objects in the personal web (i.e., heterogeneous personal information network), it is possible to identify the objects that are most relevant to specific activities or topics (e.g., work projects, vacation)~\cite{SafaviFSJWFKB20}, and then automatically organize them into groups.

\subsubsection{Role-based Embeddings}
Given the structural role-based embeddings, we can derive hard cluster assignments, which can also be used in a variety of applications.
In general, there are many applications where it would also make sense to sample a set of nodes that belong to the same structural role or a diverse set of nodes from different structural roles.
In the above, structural roles are used for sampling.
Sampling nodes from the same structural role might be useful for identifying nodes that are similar to a given node of interest.
In contrast, one possibility for sampling nodes from different structural roles is for the active learning setting.
Another important application where hard assignment of roles can be used is in selecting users with similar structural behavior for recommendation or influence maximization applications.
In online advertisement campaigns, the specific advertisement can be personalized better based on the role of a user in the network (\eg, Facebook, Yelp)~\cite{farahat2012does}.
Furthermore, a business might only be interested in targeting an individual with a certain role in the network.
Role-based embeddings and the hard assignment of roles from them are also useful in a wide variety of transfer learning tasks such as across-network classification, link prediction, and finding similar nodes in general.

\section{Conclusion} \label{sec:conc}
In this work, communities and roles are formally defined and used as a basis for analysis of the main mechanisms behind popular embedding methods for graph data.
We have described a general framework for the study of embedding methods based on whether they are community or role-based.
We have also shown formally why the mechanisms (\eg, random walks, feature diffusion) behind many of the popular embedding methods give rise to community (proximity) or role-based (structural) embeddings.
This formalization and theoretical analysis allows for a deeper understanding of the key mechanisms used by many existing embedding methods, and gives intuition for where such methods are most appropriate, but more importantly, provides intuition for how to develop better embedding methods for specific applications that may favor either communities (proximity) or roles (structural).
In addition, we discuss applications, problem settings, and data characteristics that are best for community-based or role-based embeddings.
We believe a main contribution of this work is that it allows researchers to not only gain a deeper understanding of the main mechanisms behind embedding methods (and whether they are community or role-based), but also gain insight and understanding of how to develop better embedding methods for specific applications that favor community-based or structural role-based embeddings.

\section*{Acknowledgements} \label{sec:ack}
The authors would like to thank Mark Heimann for his insightful feedback on this manuscript. 
This material is based upon work supported by the National Science Foundation under Grant No. IIS 1845491, Army Young Investigator Award No. W911NF1810397, an Adobe Digital Experience faculty research award, an Amazon faculty award, and a Google faculty award. 
Any opinions, findings, and conclusions or recommendations expressed in this material are those of the author(s) and do not necessarily reflect the views of the National Science Foundation or other funding parties. The U.S. Government is authorized to reproduce and distribute reprints for Government purposes notwithstanding any copyright notation here on.

\balance
\bibliographystyle{ACM-Reference-Format}
\bibliography{paper}

\end{document}